\documentclass[twocolumn,pra]{revtex4}
\usepackage{amssymb}
\usepackage{amsfonts}
\usepackage{amsmath}
\usepackage{amsxtra}
\usepackage{amscd}
\usepackage{amsthm}
\usepackage{graphicx}
\usepackage{epsf}
\usepackage{bbold}
\usepackage{xcolor}
\usepackage{mathalfa}
\usepackage{eucal}
\usepackage{bm}
\usepackage{array}
\newcommand{\PreserveBackslash}[1]{\let\temp=\\#1\let\\=\temp}
\newcolumntype{C}[1]{>{\PreserveBackslash\centering}p{#1}}
\newcolumntype{R}[1]{>{\PreserveBackslash\raggedleft}p{#1}}
\newcolumntype{L}[1]{>{\PreserveBackslash\raggedright}p{#1}}

\begin{document}

\title{Nonsymmorphic chiral symmetry and solitons in the Rice-Mele model}

\author{Rebecca E. J. Allen}
\affiliation{Physics Department, Lancaster University, Lancaster, LA1 4YB, UK}

\author{Holly V. Gibbons}
\affiliation{Physics Department, Lancaster University, Lancaster, LA1 4YB, UK}

\author{Alex M. Sherlock}
\affiliation{Physics Department, Lancaster University, Lancaster, LA1 4YB, UK}

\author{Harvey R. M. Stanfield}
\affiliation{Physics Department, Lancaster University, Lancaster, LA1 4YB, UK}

\author{Edward McCann}
\email{ed.mccann@lancaster.ac.uk}
\affiliation{Physics Department, Lancaster University, Lancaster, LA1 4YB, UK}

\begin{abstract}
The Rice-Mele model has two topological and spatially-inversion symmetric phases, namely the Su-Schrieffer-Heeger (SSH) phase with alternating hopping only, and the charge-density-wave (CDW) phase with alternating energies only.
The chiral symmetry of the SSH phase is robust in position space, so that it is preserved in the presence of the ends of a finite system and of textures in the alternating hopping. However, the chiral symmetry of the CDW wave phase is nonsymmorphic, resulting in a breaking of the bulk topology by an end or a texture in the alternating energies. We consider the presence of solitons (textures in position space separating two degenerate ground states) in finite systems with open boundary conditions. We identify the parameter range under which an atomically-sharp soliton in the CDW phase supports a localized state which lies within the band gap, and we calculate the expectation value $p_y$ of the nonsymmorphic chiral operator for this state, and the soliton electric charge.
As the spatial extent of the soliton increases beyond the atomic limit,
the energy level approaches zero exponentially quickly or inversely proportionally to the width, depending on microscopic details of the soliton texture. In both cases, the difference of $p_y$ from one is inversely proportional to the soliton width, while the charge is independent of the width.
We investigate the robustness of the soliton level in the presence of disorder and sample-to-sample parameter variations, comparing with a single soliton level in the SSH phase with an odd number of sites.
\end{abstract}

\maketitle

\section{Introduction}

\subsection{The Rice-Mele model}

The Rice-Mele model~\cite{ricemele82} is a one-dimensional tight-binding model with one electronic orbital per site and two sites per unit cell, with alternating onsite energies and alternating nearest-neighbor hopping. It can be considered to have two topological and spatially-inversion symmetric phases~\cite{cayssol21,fuchs21} which lie within the BDI (chiral orthogonal) classification of topological insulators~\cite{schnyder08,kitaev09,chiu16}: the Su-Schrieffer-Heeger (SSH) phase~\cite{ssh79,ssh80} has alternating hopping only, and the charge-density-wave (CDW) phase~\cite{brzezicki20,cayssol21,fuchs21} has alternating energies only.

As well as being a model of polymers~\cite{ssh79,ssh80,ricemele82,heeger88,heeger01} and of topological systems in one dimension~\cite{takayama80,kivelson83,fulga11,chen11,gangadharaiah12,pershoguba12,li14,asboth16,velasco17,bercioux17,rhim17,rhim18,liu18,munoz18,perezgonzalez19,scollon20,pletyukhov20,lin20,vanmiert20,han20,chen21,cayssol21,fuchs21}, the Rice-Mele model and its phases have been realized in engineered atomic lattices~\cite{kim12,cheon15,shim15,kim17,drost17,lee19,huda20,kiczynski22} and with cold atoms in optical lattices~\cite{atala13,przysiezna15,meier16,meier18,cooper19}. They also have analogies in higher dimensions including square lattices~\cite{seradjeh08,chamon08,liu17,vanmiert20}, graphene nanoribbons~\cite{cao17,franke18,rizzo18,groning18}, and finite stacks of rhombohedral graphite~\cite{heikkila11,xiao11,slizovskiy19,shi20}.

\begin{table*}[t]
\begin{center}
\caption{\label{tableops}Parity and chiral operations for the SSH and CDW models in $k$ space and in position space.
In $k$ space (second column), the Bloch Hamiltonian $H(k)$ is a $2 \times 2$ matrix~(\ref{rm3}), and operations are defined in terms of $2 \times 2$ Pauli matrices $\sigma_x$, $\sigma_y$, $\sigma_z$.
In position space (third column), the Hamiltonian is represented by a square matrix $H$~(\ref{rm2}) of order $J$, where $J$ is the number of atoms. $P$ is a generalization of the Pauli matrix $\sigma_x$ of order $J$~(\ref{parity}). $S_z$ is a generalization of the Pauli matrix $\sigma_z$ of order $J$~(\ref{Sz}), $T_{a/2}$ a matrix of order $J$~(\ref{ta2}) representing translation by an atomic spacing (half a lattice constant).
Time inversion and chiral-parity are symmetries of the full Rice-Mele model, and they may be combined to give charge conjugation-parity (CP) symmetry.
The final three columns show the effect of the operation on the parameters $u$, $t$, $\Delta$ of the Rice-Mele model.}
\begin{tabular}{ L{2.7cm} | C{2.3cm} | C{5cm} | C{1.0cm} | C{1.0cm} | C{1.0cm} }
\hline
operation & $k$ space & position space & $u$ & $t$ & $\Delta$ \\ 
\hline \hline
time inversion & $[ H (-k) ]^{\ast}$ & $H^{\ast}$ & $u$ & $t$ & $\Delta$  \\
\hline
SSH parity & $\sigma_x H (-k) \sigma_x$ & $P H P$ for even $J$ & $-u$ & $t$ & $\Delta$  \\  
\hline
SSH chiral & $-\sigma_z H (k) \sigma_z$ & $- S_z H S_z$ & $-u$ & $t$ & $\Delta$  \\
\hline
CDW parity & $H (-k)$ & $P H P$ for odd $J$ & $u$ & $t$ & $-\Delta$  \\  
\hline
CDW chiral & $- \sigma_y H (k) \sigma_y$ & $- S_y^{-1} H S_y$ & $u$ & $t$ & $-\Delta$  \\
 &  & where $S_y = T_{a/2} S_z$ &  &  &   \\
\hline
chiral-parity & $-\sigma_y H (-k) \sigma_y$ & $- \Gamma^{-1} H \Gamma$ & $u$ & $t$ & $\Delta$  \\
 &  & where $\Gamma =  P S_z$ for even $J$ &  &  &   \\
  &  & and $\Gamma =  P S_y$ for odd $J$ &  &  &   \\
\hline
translation by $a/2$ & $\sigma_x H (k) \sigma_x$ & $T_{a/2}^{-1} H T_{a/2}$ & $-u$ & $t$ & $-\Delta$  \\
\hline
\end{tabular}
\end{center}
\end{table*}

In position space for a system of $J$ atoms with open boundary conditions, the Rice-Mele model~\cite{ricemele82} Hamiltonian may be written as a $J \times J$ matrix in a basis of atomic orbitals,
\begin{eqnarray}
H \!=\! \begin{pmatrix}
u & t + \tfrac{1}{2}\Delta & 0 & \cdots & 0 & 0 \\
t + \tfrac{1}{2}\Delta & -u & t - \tfrac{1}{2}\Delta & \cdots &0 & 0 \\
0 & t - \tfrac{1}{2}\Delta & u & \cdots & 0 & 0 \\
\vdots & \vdots & \vdots & \vdots & \vdots & \vdots \\
0 & 0 & 0 & \cdots & u & t + \tfrac{1}{2}\Delta \\
0 & 0 & 0 & \cdots & t + \tfrac{1}{2}\Delta & -u 
\end{pmatrix} \!\! , \label{rm2}
\end{eqnarray}
where alternating onsite energies are parameterized by $u$, parameter $t$ is the mean nearest neighbor hopping, and alternating hopping is described by $\Delta$.
The alternating energies and hopping give two different atomic sites, labeled A and B. The Bloch Hamiltonian in $k$ space written in the canonical basis~\cite{cayssol21} with Bloch orbitals on A and B sites is
\begin{eqnarray}
H (k) &=&\begin{pmatrix}
u & 2 t c_k + i \Delta s_k \\
2 t c_k - i \Delta s_k & -u
\end{pmatrix} , \label{rm3} \\ 
c_k &=& \cos (ka/2) ; \qquad
s_k = \sin (ka/2) , \nonumber
\end{eqnarray}
where $a$ is the lattice constant. This has two bands with energies $E = \pm \sqrt{u^2 + 4t^2 \cos^2 (ka/2) + \Delta^2 \sin^2 (ka/2)}$. The band gap occurs at the edge of the first Brillouin zone $k = \pm \pi/a$ and has value $2 \sqrt{u^2 + \Delta^2}$.

The topological properties of the SSH phase ($u=0$) have been studied at length~\cite{ssh79,ssh80,takayama80,chen11,gangadharaiah12,li14,asboth16,rhim18,liu18,munoz18,perezgonzalez19,scollon20,vanmiert20,han20,chen21,cayssol21,fuchs21}, and ends and solitons in the $\Delta$ texture (domain walls in position space separating two degenerate ground states) conserve the chiral symmetry. The CDW phase ($\Delta=0$) is less well studied~\cite{brzezicki20,cayssol21,fuchs21}; the chiral symmetry is nonsymmorphic~\cite{shiozaki15,zhao16,brzezicki20,han20} so that solitons in $u$ and ends in a finite system break the chiral symmetry.
In this paper, we focus on the properties of non-topological solitons in the CDW phase, and consider whether they are robust to disorder or sample-to-sample variations in parameter values.

In the remainder of this Introduction, we review the symmetry properties of the Rice-Mele model, describing the representation of the nonsymmorphic chiral operator in position space. Then, we review the Jackiw-Rebbi mechanism~\cite{jackiwrebbi76} which, in the continuum limit, predicts the existence of solitons which preserve chiral symmetry and support localized states at zero energy. In Section~\ref{s:atomic}, we discuss the properties of atomically-sharp solitons (where the texture is essentially a step function) in the CDW wave phase in a finite system. We employ numerical calculations as well as perturbation theory for weak hopping ($|t| \ll |u|$) to show that these solitons support a localized state with energy within the band gap for a wide range of parameter values but that, by tuning the hopping $t$ to very large values determined by the system size, the soliton level will eventually merge with the bulk states. The soliton state is characterized by the expectation value of the nonsymmorphic chiral operator, $p_y$, which is a generalization of electric polarization.

Section~\ref{s:smooth} describes the properties of a soliton with a spatially-smooth texture of width $\xi$ greater than the lattice constant. As the width $\xi$ increases, the soliton energy approaches zero exponentially quickly or inversely proportionally to $\xi$, depending on microscopic details of the soliton texture~\cite{brzezicki20}. In both cases, the difference of $p_y$ from one is only inversely proportional to $\xi$, i.e. even a level at zero energy is not topological in a finite system.
In Section~\ref{s:charge} we show numerically that the electric charge of an atomically-sharp soliton isn't half integer in the CDW phase, unlike the SSH phase~\cite{jackiwrebbi76,heeger88} (for spinless electrons at half filling), but dependent on the ratio $u / t$ of the parameters~\cite{brzezicki20}.
We find that the charge is independent of soliton width $\xi$, so a smooth soliton has the same charge as an atomically-sharp one, as determined by $u/t$, even though its energy and $p_y$ value are different.
In Section~\ref{s:disorder}, we discuss solitons in disordered systems~\cite{inui94,perezgonzalez19,scollon20}. For a Hamiltonian to satisfy the nonsymmorphic chiral symmetry in position space, its form is highly restricted, and its parameters must be uniform across the sample. This means that spatial disorder will break the chiral symmetry, but that some types of sample-to-sample variations (e.g. induced by a gate potential) will conserve the chirality. Using numerical calculations, we study the effects of disorder and sample-to-sample variations on a soliton in the CDW phase, comparing with the properties of a soliton in the SSH phase.

\subsection{Symmetries of the Rice-Mele model}

The symmetries of the Rice-Mele model are summarized in Table~\ref{tableops}. The model satisfies time-inversion symmetry~\cite{cayssol21} so that the position space Hamiltonian~(\ref{rm2}) is real. For the SSH phase, a center of inversion is mid-bond and spatial-inversion symmetry (parity) involves swapping A and B sites, as described by $\sigma_x$ in $k$ space~\cite{cayssol21}. In position space, this only holds for an even number of atoms $J$ and is represented by a matrix $P$ of order $J$,
\begin{eqnarray}
P = \begin{pmatrix}
0 & 0 & 0 & \cdots & 0 & 1 \\
0 & 0 & 0 & \cdots & 1 & 0 \\
\vdots & \vdots & \vdots & \vdots & \vdots & \vdots \\
0 & 0 & 1 & \cdots & 0 & 0 \\
0 & 1 & 0 & \cdots & 0 & 0 \\
1 & 0 & 0 & \cdots & 0 & 0 
\end{pmatrix} \!\! . \label{parity}
\end{eqnarray}
For the CDW phase, a center of inversion is an atomic site so parity doesn't involve swapping A and B sites~\cite{cayssol21}. In position space, parity is still represented by matrix $P$ of order $J$ as in Eq.~(\ref{parity}), but this holds for odd $J$ only.

For the SSH phase, chiral symmetry is represented by $\sigma_z$ in $k$ space, and this may be represented in position space as a matrix $S_z$ of order $J$,
\begin{eqnarray}
S_z = \begin{pmatrix}
1 & 0 & 0 & \cdots & 0 & 0 \\
0 & -1 & 0 & \cdots & 0 & 0 \\
0 & 0 & 1 & \cdots & 0 & 0 \\
\vdots & \vdots & \vdots & \vdots & \vdots & \vdots \\
0 & 0 & 0 & \cdots & 1 & 0 \\
0 & 0 & 0 & \cdots & 0 & -1 \\
\end{pmatrix} . \label{Sz}
\end{eqnarray}
Note that $S_z$ may be generalized to an odd number of atoms with termination of $1$ (instead of $-1$) at the bottom right corner~\cite{chen21}.
The chiral symmetry~(\ref{Sz}) for the SSH model is extremely robust because $S_z$ is diagonal in the sublattice space. This means that the chiral symmetry holds even in the presence of a position-dependent texture in $\Delta$.

Note that the position space Hamiltonian~(\ref{rm2}) is real so the SSH phase satisfies time-inversion symmetry~\cite{cayssol21}. With chiral symmetry, this places the SSH phase in the BDI (chiral orthogonal) classification of topological insulators~\cite{schnyder08,kitaev09,chiu16}. If the hopping parameters acquired a complex phase, the SSH phase would not satisfy time-inversion symmetry and would lie in the AIII (chiral unitary) symmetry class~\cite{velasco17}, but this is not a case we consider in this paper.

Although chiral symmetry of the CDW phase is simply represented by $\sigma_y$ in $k$ space, this is nonsymmorphic~\cite{shiozaki15,zhao16,brzezicki20,han20}. In position space, it can be represented as $S_y = T_{a/2} S_z$ which is a matrix product of $S_z$~(\ref{Sz}) with $T_{a/2}$ describing translation by an atomic spacing $a/2$:
\begin{eqnarray}
T_{a/2} = \begin{pmatrix}
0 & 1 & 0 & 0 & \cdots & 0 & 0 \\
0 & 0 & 1 & 0 & \cdots & 0 & 0 \\
0 & 0 & 0 & 1 & \cdots & 0 & 0 \\
0 & 0 & 0 & 0 & \cdots & 0 & 0 \\
\vdots & \vdots & \vdots & \vdots & \vdots & \vdots & \vdots \\
0 & 0 & 0 & 0 & \cdots & 0 & 1 \\
1 & 0 & 0 & 0 & \cdots & 0 & 0 \\
\end{pmatrix} , \label{ta2}
\end{eqnarray}
\begin{eqnarray}
S_{y} = \begin{pmatrix}
0 & -1 & 0 & 0 & \cdots & 0 & 0 \\
0 & 0 & 1 & 0 & \cdots & 0 & 0 \\
0 & 0 & 0 & -1 & \cdots & 0 & 0 \\
0 & 0 & 0 & 0 & \cdots & 0 & 0 \\
\vdots & \vdots & \vdots & \vdots & \vdots & \vdots & \vdots \\
0 & 0 & 0 & 0 & \cdots & 0 & -1 \\
1 & 0 & 0 & 0 & \cdots & 0 & 0 \\
\end{pmatrix} , \label{Sy}
\end{eqnarray}
where $S_y$ is written for even $J$.
In contrast to $S_z$, chiral symmetry in the CDW phase is fragile in position space, and it is violated by an end or a texture in the alternating energies.
Note that a chiral symmetry which is a combination of sublattice symmetry and a shift of the energy spectrum (i.e. a shift proportional to the identity matrix) has recently been discussed in the context of non-Hermitian systems~\cite{kawasaki22}.

In Table~\ref{tableops}, we also include chiral-parity~\cite{arkinstall17,brzezicki20} which is a symmetry of the Rice-Mele model. It may be combined with time-inversion to give charge conjugation-parity (CP) symmetry, guaranteeing electron-hole symmetry of the electronic spectrum.

\subsection{Jackiw-Rebbi mechanism}

Although the chiral symmetry $S_y$ of the CDW phase tight-binding model~(\ref{Sy}) is fragile, there are levels at zero energy localized on a soliton in the continuum limit, as described by the Jackiw-Rebbi mechanism~\cite{jackiwrebbi76}.
The continuum Hamiltonian is obtained by substituting $k \rightarrow - (\pi / a) + \hat{p}/\hbar$~\cite{continuumnote} in $H(k)$, Eq.~(\ref{rm3}), where $\hat{p}$ is the momentum operator,
\begin{eqnarray*}
H = v \hat{p} \sigma_x + \Delta (x) \sigma_y + u(x) \sigma_z ,
\end{eqnarray*}
where the velocity is $v = a t /\hbar$.
In the SSH phase, $u(x) = 0$, we consider a soliton profile of the staggered hoppings $\Delta (x)$ centred on $x=0$ with limits given by
\begin{eqnarray}
\lim_{x \rightarrow - \infty} \Delta (x) = - s\Delta_0 ; \qquad
\lim_{x \rightarrow \infty} \Delta (x) = s\Delta_0  ,
\end{eqnarray}
for $\Delta _0 > 0$ with parameter $s = \pm 1$ describing two different textures.
Then, there is a single localized state for each texture with energy $E = 0$~\cite{jackiwrebbi76,ssh79,ssh80,cayssol21}, and (unnormalized) wave functions given by
\begin{eqnarray}
\psi_{s} (x) = e^{- \frac{s}{\hbar v} \int_0^x \Delta(x^{\prime}) dx^{\prime}}
\begin{pmatrix}
(1-s)/2 \\
(1+s)/2
\end{pmatrix} .
\end{eqnarray}

For the CDW phase, $\Delta (x) = 0$, we consider a soliton profile of the onsite energies $u(x)$ centred on $x=0$ with limits given by
\begin{eqnarray}
\lim_{x \rightarrow - \infty} u(x) = - su_0 ; \qquad
\lim_{x \rightarrow \infty} u(x) = su_0  ,
\end{eqnarray}
for $u_0 > 0$ with parameter $s = \pm 1$ describing two different textures.
Then, there is a single state with energy $E = 0$~\cite{kivelson83} and (unnormalized) wave function given by
\begin{eqnarray}
\psi_s (x) = e^{- \frac{s}{\hbar v} \int_0^x u(x^{\prime}) dx^{\prime}}
\begin{pmatrix}
1 \\
is
\end{pmatrix} .
\end{eqnarray}
Thus, a smooth soliton in the continuum limit should support a zero-energy state with topological properties~\cite{brzezicki20}. It is the aim of this paper to model the properties of solitons in finite systems where the chiral symmetry of the CDW phase is broken. We begin by considering atomically-sharp solitons and describing the parameter values for which they support a localized state with energy within the bulk band gap.

\section{Atomically-sharp solitons}\label{s:atomic}

\subsection{Qualitative picture for weak hopping}

\begin{figure}[t]
\includegraphics[scale=0.35]{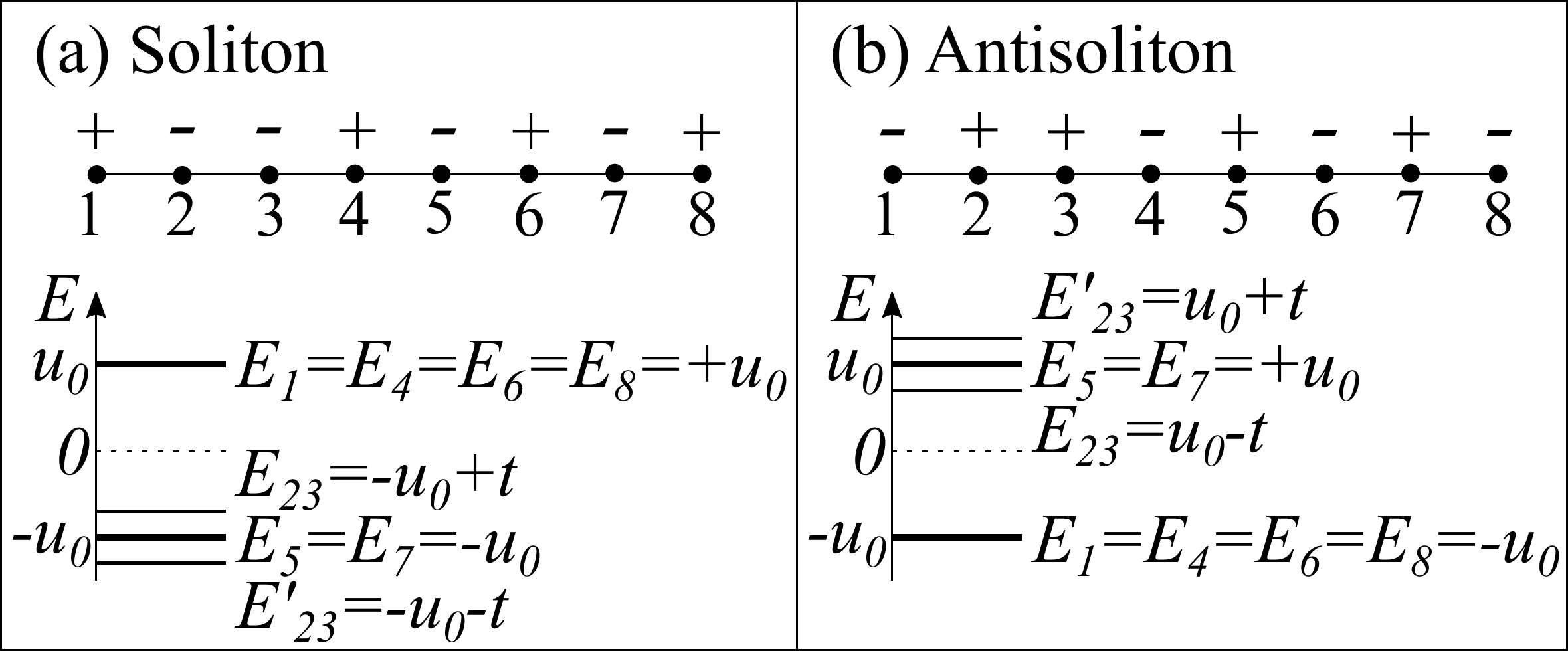}
\caption{Atomically-sharp domain walls in the CDW phase, shown schematically for a finite system with open boundary conditions and $J=8$ sites, where numbers $1,2,\ldots J$ label sites. Straight lines indicate nearest-neighbor hopping $t$ and $\pm$ symbols indicate alternating onsite energies $u_0$ and $-u_0$, with $u_0>0$. (a) a soliton consisting of two consecutive onsite energies $-u_0$ on sites 2 and 3. For hopping $t=0$, energy levels are degenerate at $E = u_0$ or $E = -u_0$. To first order in degenerate perturbation theory in $t$, the levels on sites 2 and 3 are split as $E_{23}^{\prime} = -u_0-t$ and $E_{23} = -u_0+t$. The latter lies within the band gap $-u_0 < E < u_0$, and we refer to it as a soliton because it generally lies at negative energy. (b) an antisoliton consisting of two adjacent onsite energies $u_0$ on sites 2 and 3. To first order in degenerate perturbation theory in $t$, the levels on sites 2 and 3 are split as $E_{23}^{\prime} = u_0+t$ and $E_{23} = u_0-t$. The latter lies within the band gap $-u_0 < E < u_0$, and we refer to it as an antisoliton because it generally lies at positive energy.
}\label{dptplot}
\end{figure}

We consider atomically-sharp domain walls in the CDW phase, Fig.~\ref{dptplot}.
As an illustrative example, Figure~\ref{dptplot}(a) shows a finite system with open boundary conditions and $J=8$ atoms, and with a soliton consisting of two consecutive onsite energies $-u_0$ on sites 2 and 3.
We use $u_0$ (where $u_0 \geq 0$) to denote the magnitude of the staggered onsite energies $u$ at the ends of the system (away from the soliton). In the case of an atomically-sharp soliton, every atom in the system has an onsite energy of $\pm u_0$. We also assume $t \geq 0$.

To describe the origin of a state within the band gap and localized on the soliton, we consider the regime of weak nearest-neighbor hopping $t < u_0$. At $t=0$, all states are localized on independent atoms and have energies determined by the onsite energies, namely $E = +u_0$ or $E = - u_0$, and they are highly degenerate. Using degenerate perturbation theory, to first order in $t$, hopping $t$ hybridizes the two degenerate orbitals immediately adjacent to the soliton [sites 2 and 3 in Figure~\ref{dptplot}(a)]. In the $2 \times 2$ space of these two atoms, with $\psi_2^T = \begin{pmatrix} 1 & 0 \end{pmatrix}$ and $\psi_3^T = \begin{pmatrix} 0 & 1 \end{pmatrix}$, the interaction is $t \sigma_x$. It results in splitting of these two levels, as $E_{23}^{\prime} = -u_0-t$ and $E_{23} = -u_0+t$. The latter lies within the band gap $-u_0 < E < u_0$, and we refer to it as a soliton because it generally lies at negative energy. Note that it corresponds to the antibonding (higher-energy) state between sites 2 and 3, i.e. $\psi_{23}^T = \begin{pmatrix} 1 & 1 \end{pmatrix} / \sqrt{2}$. To second order perturbation in $t$, we find its energy eigenvalue $E_{\mathrm{sol}}$ is
\begin{eqnarray}
E_{\mathrm{sol}} = - u_0 + t - \frac{t^2}{2u_0} , \label{esol}
\end{eqnarray}
where the second order term describes interaction with the two closest sites with opposite energy [sites 1 and 4 in Figure~\ref{dptplot}(a)]. We note that this expression is independent of the length of the system or the position of the soliton (assuming a soliton always occurs between unit cells) because it only involves orbitals on the four sites near the soliton.

Figure~\ref{dptplot}(b) shows an antisoliton which consists of two consecutive onsite energies $u_0$ on sites 2 and 3; we name this an antisoliton because it is generally at positive energy. Note that it arises from bonding between the two sites, $\psi_{23}^T = \begin{pmatrix} 1 & -1 \end{pmatrix} / \sqrt{2}$. To second order in perturbation in $t$, we find its energy eigenvalue is $E_{\mathrm{antisol}} = - E_{\mathrm{sol}}$, with $E_{\mathrm{sol}}$ given in Eq.~(\ref{esol}).

We determine expectation values of the chiral operators $S_z$ and $S_y$ for the soliton state which are generalized versions of electric polarization,
\begin{eqnarray}
p_z &=& \langle \psi | S_z | \psi \rangle , \\ 
p_y &=& \langle \psi | S_y | \psi \rangle , \label{pydef}
\end{eqnarray}
where $S_y = T_{a/2} S_z$.
Using degenerate perturbation theory for small $t$, it is sufficient to consider only four atoms in the vicinity of a soliton [e.g. sites $1$-$4$ in Figure~\ref{dptplot}(a)]. To first order in $t$, the soliton state is $\psi^{T} = \begin{pmatrix}
- \tau & 1 & 1 & - \tau \end{pmatrix} / \sqrt{2(1+\tau^2)}$ for $\tau = t / (2u_0)$. This yields $p_z^{\mathrm{sol}} = 0$ and
\begin{eqnarray}
p_y^{\mathrm{sol}} = \frac{(1 + \tau)^2}{2(1+\tau^2)} , \qquad \tau = \frac{t}{2u_0} . \label{pya}
\end{eqnarray}
This predicts $p_y^{\mathrm{sol}} = 1/2$ for $t = 0$ and $p_y^{\mathrm{sol}} > 1/2$ for $t > 0$.
For the antisoliton, $p_y^{\mathrm{antisol}} = - p_y^{\mathrm{sol}}$.

\subsection{Numerical results}

Energy eigenvalues $E_n$ and eigenstates $\psi_n$, $n = 1,2,\ldots , J$, are obtained by numerical diagonalization of the position space Hamiltonian~(\ref{rm2}) with a texture in the onsite energies $u$ (and $\Delta = 0$ for the CDW phase).
The soliton state has index $n = J/2$ for even $J$.
The density of states per unit energy $g (E)$ is determined numerically by approximation using a Lorentzian with a finite width~$\delta$,
\begin{eqnarray}
g (E) = \frac{1}{\pi} \sum_n \frac{\delta}{(E - E_n)^2 + \delta^2} . \label{dos}
\end{eqnarray}

\begin{figure}[t]
\includegraphics[scale=0.36]{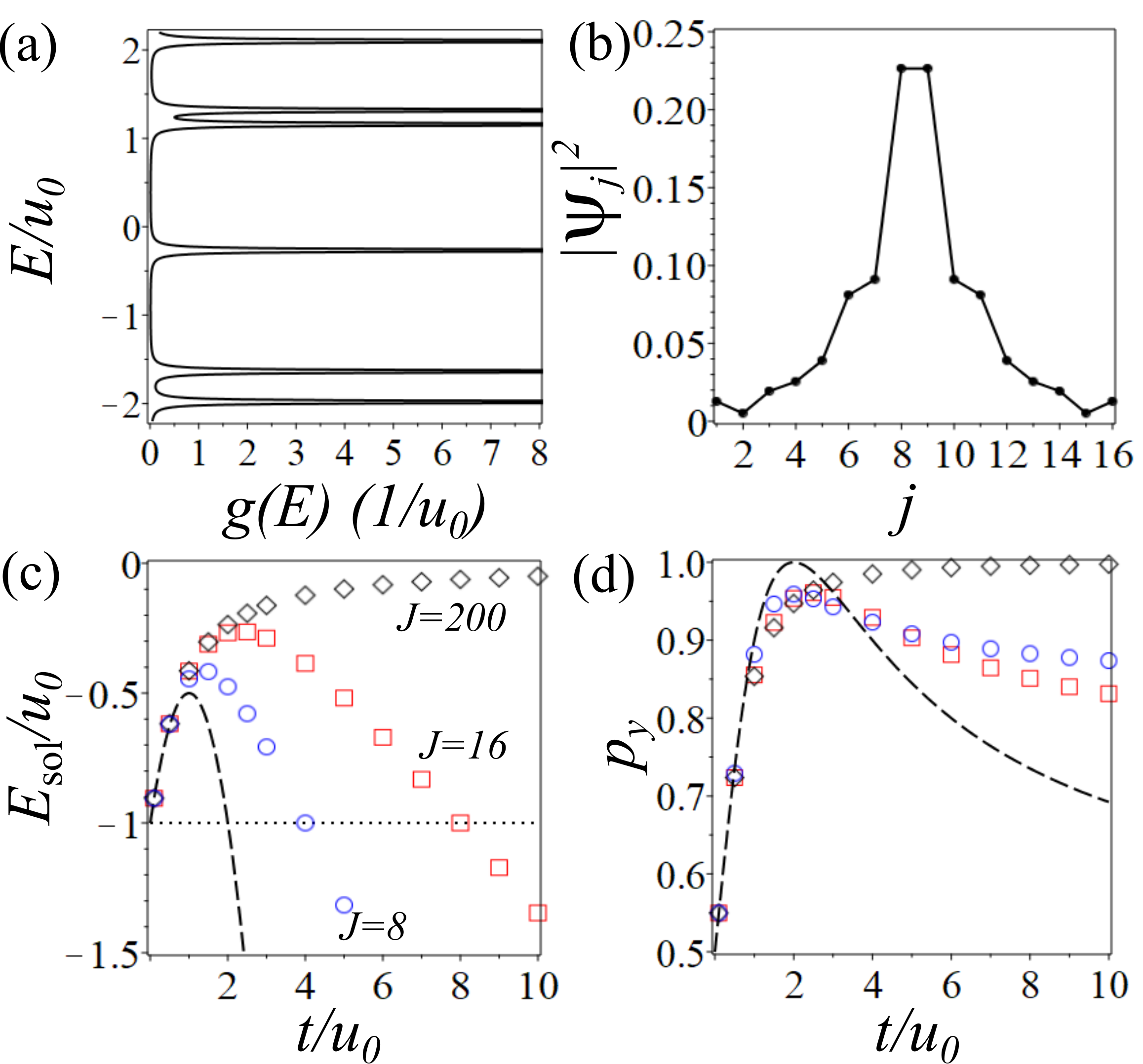}
\caption{A single atomically-sharp soliton in the CDW phase, at the center of a finite system with open boundary conditions.
(a) shows the density of states determined numerically for a system with $J = 16$ atoms using Eq.~(\ref{dos}) with broadening $\delta = 0.005u_0$. (b) is the probability density $|\psi_j|^2$ per site $j = 1,2,\ldots,16$ for the energy level localized on the soliton (with energy $E_{\mathrm{sol}} = -0.268u_0$). In (a) and (b), $t=2.0u_0$. (c) shows the energy eigenvalue $E_{\mathrm{sol}}$ plotted as a function of the ratio $t/u_0$ of the hopping to the alternating onsite energy, where the dotted line indicates the bulk band edge $E = - u_0$. (d) shows the polarization $p_y$ of the soliton eigenstate, where $p_y$ measures the chiral symmetry of the CDW phase~(\ref{pydef}). In (c) and (d), black diamonds are numerical data points for a system with $J = 200$ atoms, red squares are numerical data points for $J = 16$ atoms, and blue circles are numerical data points for $J = 8$ atoms. Dashed lines represent the predictions of degenerate perturbation theory for $t \ll u_0$, namely Eq.~(\ref{esol}) for the soliton energy and Eq.~(\ref{pya}) for the polarization $p_y$.
}\label{cdw2plot}
\end{figure}

We begin by discussing when a single atomically-sharp soliton results in a localized state with an energy level within the bulk band gap.
We consider a system with an even number of atoms, $J$, where $J = 2N$ with $N$ unit cells.
Fig.~\ref{cdw2plot} shows a single atomically-sharp soliton in the CDW phase located at the center of a finite system with open boundary conditions (there are an even number of unit cells in total).
Fig.~\ref{cdw2plot}(a) shows the density of states determined numerically for a system with $J = 16$ atoms using Eq.~(\ref{dos}) with broadening $\delta = 0.005u_0$ and $t/u_0=2.0$. An energy level can be observed within the bulk band gap $-u_0 < E < u_0$ at $E_{\mathrm{sol}} = -0.268u_0$. Fig.~\ref{cdw2plot}(b) plots the probability density $|\psi_j|^2$ per site $j = 1,2,\ldots,16$ for the state corresponding to this level, showing that the state is localized at the soliton.

Fig.~\ref{cdw2plot}(c) shows the energy eigenvalue $E_{\mathrm{sol}}$ of a single state plotted as a function of the ratio $t/u_0$ of the hopping to the alternating onsite energy, for different system sizes. The horizontal dotted line shows the band edge $E = - u_0$, and we find that there is a single level within the gap, but below zero energy, $-u_0 < E_{\mathrm{sol}} < 0$ for $0 < t / u_0 < J/2$ where $J$ is the number of atoms. For small $t/u_0$, there is agreement with the prediction of perturbation theory Eq.~(\ref{esol}) (dashed line). For large $t/u_0$, the energy is exactly $E_{\mathrm{sol}} = -u_0$ for $t/u_0=J/2$. This can be shown analytically, as in Appendix~\ref{a:bandedge}. Fig.~\ref{cdw2plot}(c) shows this energy explicity for $J=8$ (blue circles) and for $J=16$ (red squares).

The polarization $p_y$ is plotted in Fig.~\ref{cdw2plot}(d) and, for small $t/u_0$, there is agreement with the prediction of perturbation theory Eq.~(\ref{pya}) (dashed line).
The energy level $E_{\mathrm{sol}}$ is within the band gap for a wide range of parameters, although not necessarily near zero energy. It can approach arbitrarily close to zero, with polarization $p_y$ arbitrarily close to one, for a system that is sufficiently long $J \gg 1$ and with tuned parameters, typically $t > u_0$ (black diamonds in Fig.~\ref{cdw2plot} show numerical data for $J=200$). 
However, the existence of the energy at the band edge $E_{\mathrm{sol}} = -u_0$ for $t/u_0=J/2$ explicitly demonstrates that a finite system breaks the bulk topology: parameter $t$, which conserves the bulk chiral symmetry, can be tuned to a high value, moving the level away from zero energy and out of the band gap at $t/u_0=J/2$.

\begin{figure}[t]
\includegraphics[scale=0.36]{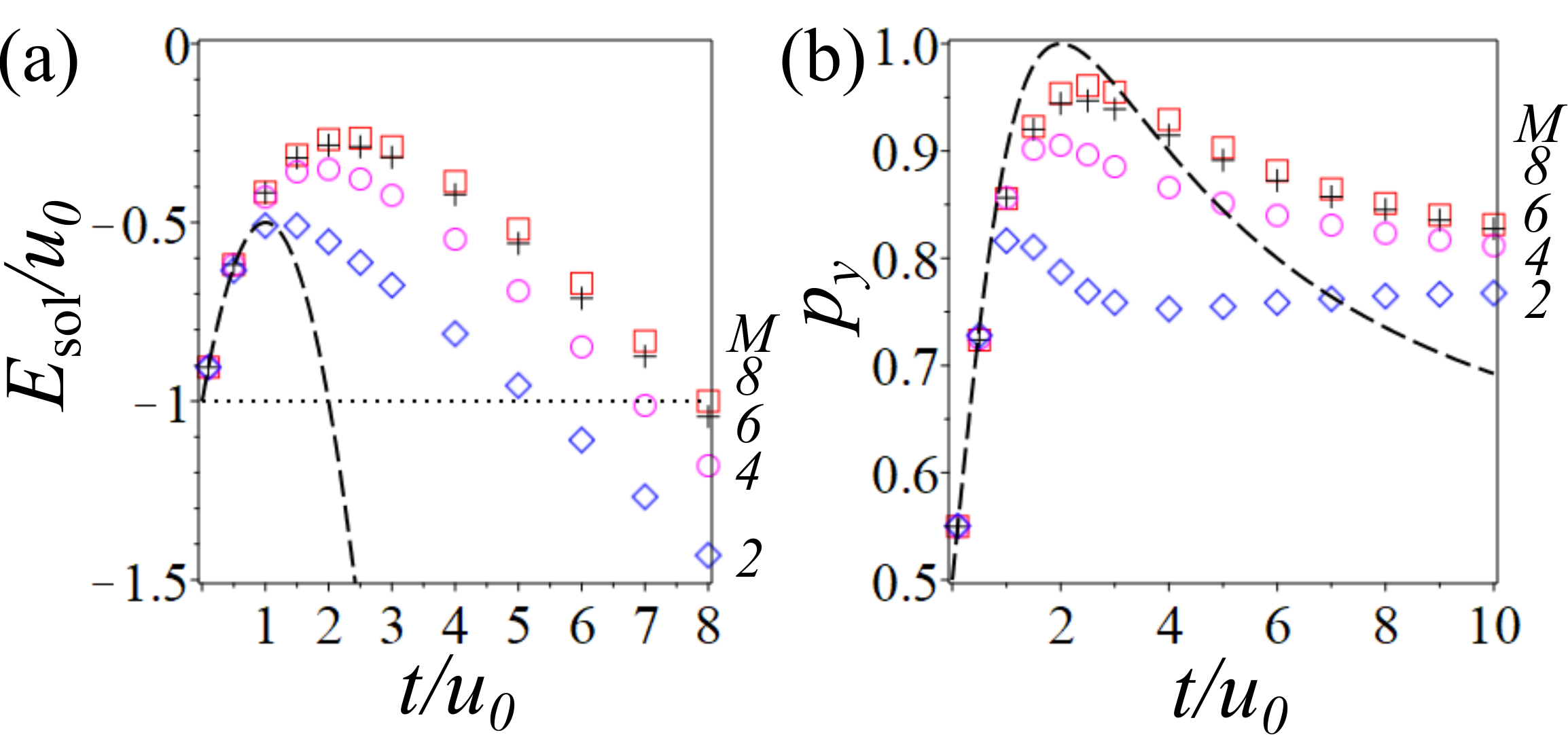}
\caption{A single atomically-sharp soliton in the CDW phase, in a finite system of $J=16$ atoms with open boundary conditions. The soliton is placed in different positions, with numerical data showing the soliton at a distance from the closest end of $M = 2$ atoms (blue diamonds), $4$ atoms (magenta circles), $6$ atoms (black crosses), and $8$ atoms (red squares).
(a) shows the energy eigenvalue of a single soliton $E_{\mathrm{sol}}$ plotted as a function of the ratio $t/u_0$ of the hopping to the alternating onsite energy, where the dotted line indicates the bulk band edge $E = - u_0$. (b) shows the polarization $p_y$ of the soliton eigenstate, where $p_y$ measures the chiral symmetry of the CDW phase~(\ref{pydef}). In both plots, dashed lines represent the predictions of degenerate perturbation theory for $t \ll u_0$, namely Eq.~(\ref{esol}) for the soliton energy and Eq.~(\ref{pya}) for the polarization $p_y$.
}\label{cdw3plot}
\end{figure}

Fig.~\ref{cdw3plot} shows a single atomically-sharp soliton in the CDW phase located at different positions in a finite system of $J=16$ atoms with open boundary conditions. Again, the horizontal dotted line shows the band edge $E = - u_0$, and we find that there is a single state within the gap, but below zero energy, $-u_0 < E_{\mathrm{sol}} < 0$ for a range of $t$ values. At the very least, this range is $0 < t/u_0 < M$ where $M$ is the number of atoms between the soliton and the closest end of the system.

For an atomically-sharp soliton, the energy level $E_{\mathrm{sol}}$ can approach zero and the polarization $p_y$ can approach one for a long system by increasing the ratio $t/u_0$.
However, increasing the ratio $t/u_0$ also increases the total band width ($4t$) as compared to the band gap ($2u_0$). An alternative way to tune the properties of the soliton is to make it spatially smooth with a characteristic width greater than the lattice constant.

\begin{figure}[t]
\includegraphics[scale=0.36]{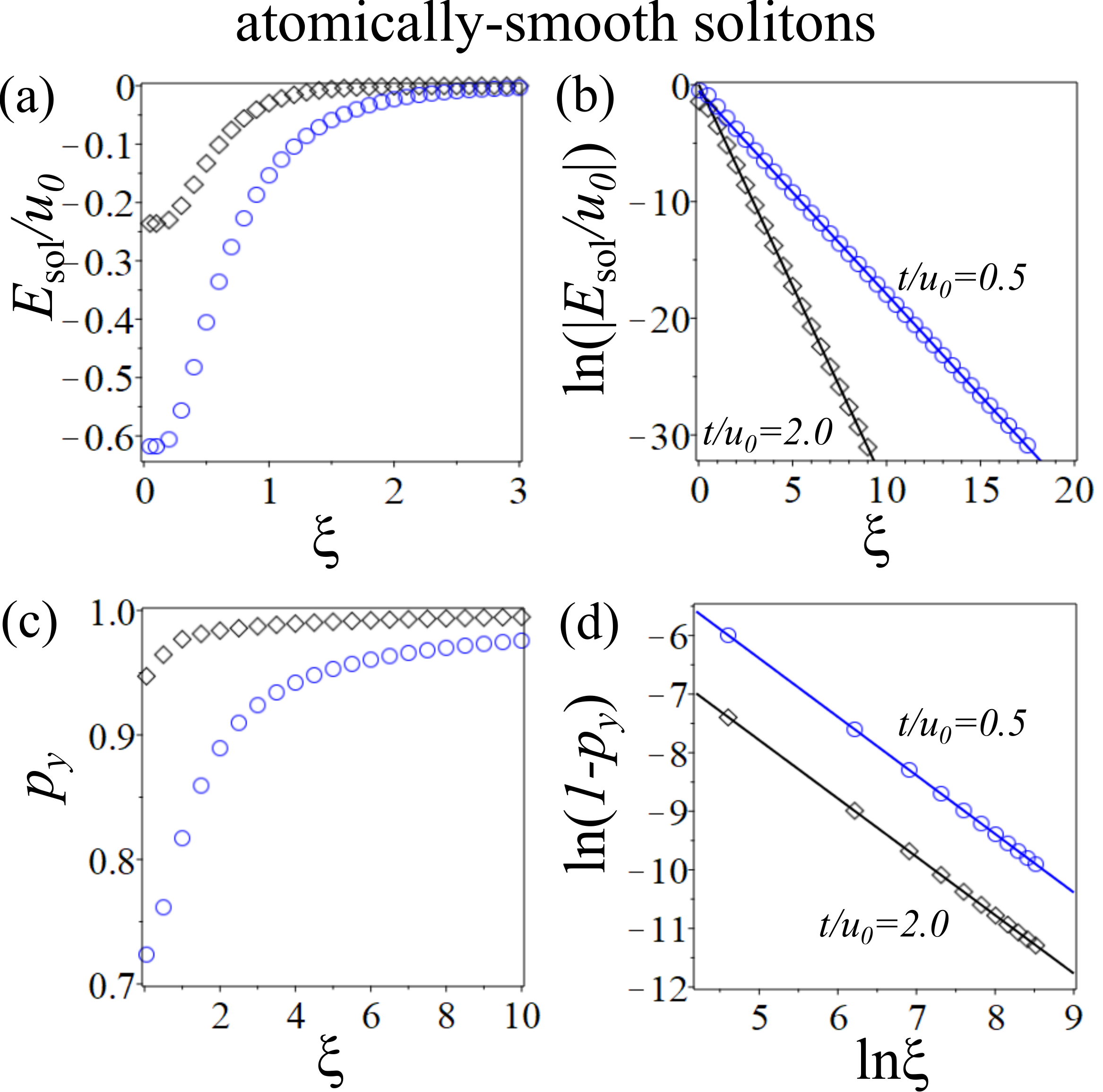}
\caption{Dependence in the CDW phase on soliton width $\xi$, for a single atomically-smooth soliton~(\ref{smoothsol}) at the center of a finite system with open boundary conditions and $J = 5000$ atoms. The width $\xi$ is dimensionless (the physical width in units of the atomic spacing $a/2$). In all plots, black diamonds are numerical data points for $t/u_0=2.0$, and blue circles are for $t/u_0 = 0.5$. 
(a) shows the soliton energy level $E_{\mathrm{sol}}$, (b) shows $\ln (|E_{\mathrm{sol}} / u_0|)$ versus $\xi$ with linear fits (solid lines). Fitting is done with data up to the point where $E_{\mathrm{sol}}$ is zero within numerical precision ($2.0 \leq \xi \leq 9.0$ for $t/u_0=2.0$ and $4.0 \leq \xi \leq 17.5$ for $t/u_0=0.5$).
(c) shows the polarization $p_y$, (d) shows a log-log plot with $\ln (1-p_y)$ versus $\ln \xi$ and linear fits (solid lines). Fitting is done with data for $100 \leq \xi \leq 5000$.
}\label{cdw4plot}
\end{figure}

\section{Smooth solitons}\label{s:smooth}

We determine the properties of spatially-smooth solitons, generalizing the atomically-sharp domain wall in Figure~\ref{dptplot}(a). There are two ways~\cite{brzezicki20} to model smooth solitons in the CDW phase: (i) atomically-smooth solitons where the energies vary smoothly on the atomic scale so that the magnitude of onsite energies of A and B sites within the same unit cell are slightly different, and (ii) unit-cell-smooth solitons where the energies within the unit cell (on A and B sites) have the same magnitude.
Ref~\cite{brzezicki20} showed that the energy $E_{\mathrm{sol}}$ of the former depends exponentially on soliton width whereas the energy $E_{\mathrm{sol}}$ of the latter varies inversely proportionally to the width. Here we consider both types of soliton, including the dependence of $E_{\mathrm{sol}}$ on parameters (i.e. the ratio $t/u_0$) and the behavior of the polarization $p_y$.

\subsection{Atomically-smooth solitons}

To model an atomically-smooth soliton, we implement onsite energies $u_j$ with site index $j = 1,2,\ldots J$ as
\begin{eqnarray}
u_j = (-1)^{j} u_0 \tanh \left( \frac{j - j_0}{\xi} \right) , \label{smoothsol}
\end{eqnarray}
where $u_0$ is the magnitude at infinity, and $\xi$ is the width in dimensionless units written as the physical width divided by the atomic spacing ($a/2$). For domain walls centred between unit cells, the center $j_0$ should be an even number plus $1/2$, e.g. for a centre between sites 2 and 3 as in Figure~\ref{dptplot}(a), then $j_0 = 5/2$. The energy profile for an antisoliton is the same as in Eq.~(\ref{smoothsol}) but with an additional minus sign.

\begin{figure}[t]
\includegraphics[scale=0.36]{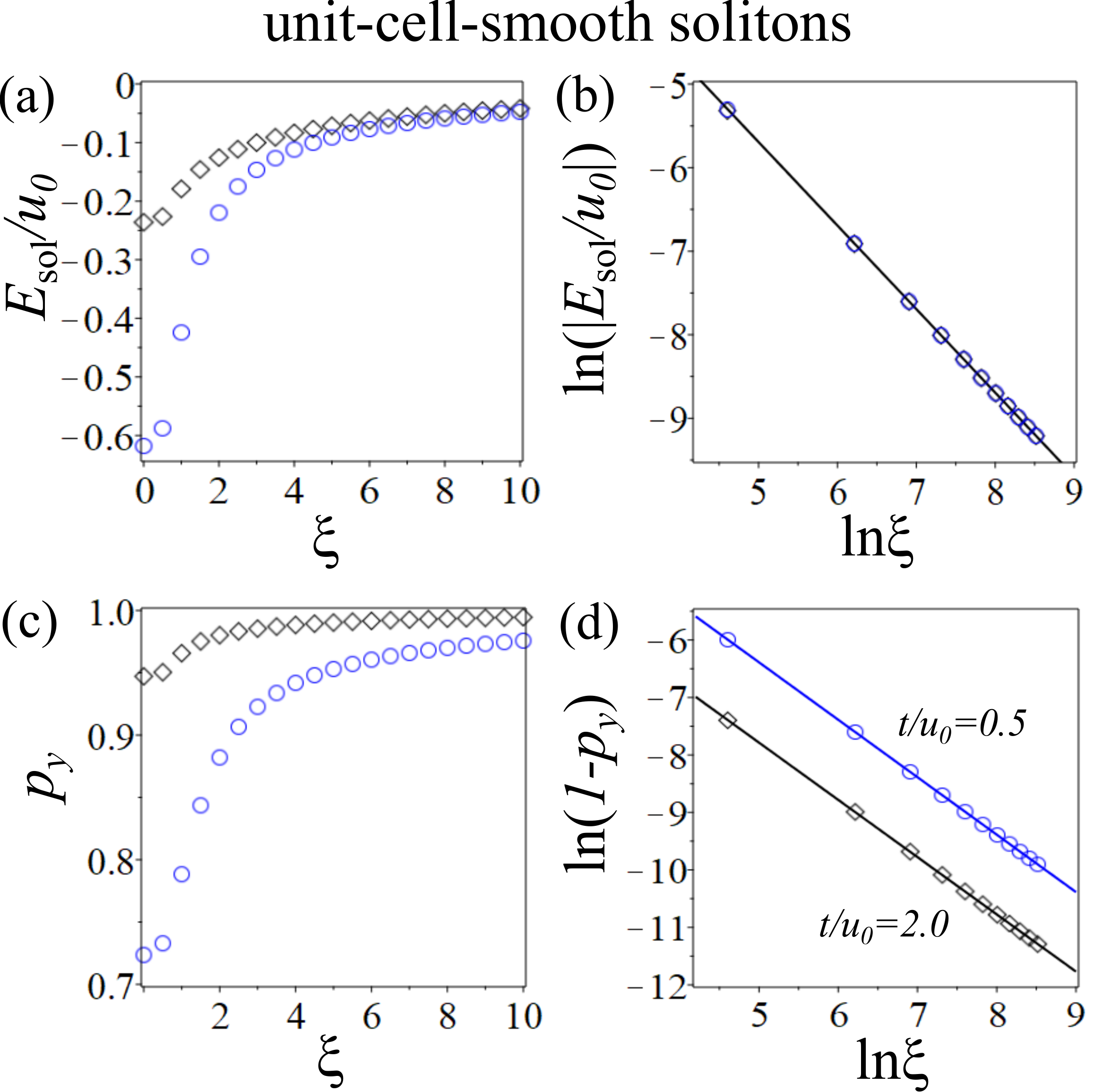}
\caption{Dependence in the CDW phase on soliton width $\xi$, for a single unit-cell-smooth soliton~(\ref{unitsmoothsol}) at the center of a finite system with open boundary conditions and $J = 5000$ atoms. The width $\xi$ is dimensionless (the physical width in units of the atomic spacing $a/2$). In all plots, black diamonds are numerical data points for $t/u_0=2.0$, and blue circles are for $t/u_0 = 0.5$. 
(a) shows the soliton energy level $E_{\mathrm{sol}}$, (b) shows a log-log plot of $\ln (|E_{\mathrm{sol}} / u_0|)$ versus $\ln \xi$ with the solid line showing $E_{\mathrm{sol}}/u_0 = - 1/(2\xi)$~(\ref{eunitsmooth}). Note that numerical data points (for $t/u_0=0.5$ and for $t/u_0 = 2.0$) coincide.
(c) shows the polarization $p_y$, (d) shows a log-log plot with $\ln (1-p_y)$ versus $\ln \xi$ and linear fits (solid lines). Fitting is done with data for $100 \leq \xi \leq 5000$.
}\label{unitcellsmoothplot}
\end{figure}

Fig.~\ref{cdw4plot} shows the dependence of $E_{\mathrm{sol}}$ and $p_y$ on soliton width $\xi$ for a single soliton~(\ref{smoothsol}) at the center of a finite system with open boundary conditions and $J = 5000$ atoms. Fig.~\ref{cdw4plot}(a) and (b) show that $E_{\mathrm{sol}}$ approaches zero exponentially quickly~\cite{brzezicki20} with $\xi$,
\begin{eqnarray}
E_{\mathrm{sol}} \propto - e^{- \xi / \ell} , \label{eatomicsmooth}
\end{eqnarray}
with a $\xi$-independent parameter $\ell$.
However, Fig.~\ref{cdw4plot}(c) and (d) show that $p_y$ approaches one much more slowly. We fit $\ln (1 - p_y )$ versus $\ln \xi$ to a straight line $y = mx +c$ with data in the range $500 \leq \xi \leq 5000$. Repeating this for different data sets in the range $0.1 \leq t/u_0 \leq 10.0$ yields $m = - 0.998 \pm 0.001$, so we deduce that the difference $1 - p_y$ is inversely proportional to $\xi$,
 \begin{eqnarray}
1 - p_y \propto \frac{1}{\xi} . \label{pysmooth}
\end{eqnarray}
Thus, it is possible to have a zero energy state (within numerical precision) that breaks the bulk topology as indicated by non-integer $p_y$, even for $\xi \gg 1$.

\subsection{Unit-cell-smooth solitons}

To model a unit-cell-smooth soliton, we implement onsite energies $u_j$ with site index $j = 1,2,\ldots J$ as
\begin{eqnarray}
u_j = \pm u_0 \tanh \left( \frac{j \mp 1/2 - j_0}{\xi} \right) , \label{unitsmoothsol}
\end{eqnarray}
where the plus (minus) sign is for the B (A) atom in the unit cell, $u_0$ is the magnitude at infinity, and $\xi$ is the width in dimensionless units written as the physical width divided by the atomic spacing ($a/2$). For domain walls centred between unit cells, the center $j_0$ should be an even number plus $1/2$. The energy profile for an antisoliton is the same as in Eq.~(\ref{unitsmoothsol}) but with an additional minus sign.

Fig.~\ref{unitcellsmoothplot} shows the dependence of $E_{\mathrm{sol}}$ and $p_y$ on soliton width $\xi$ for a single soliton~(\ref{unitsmoothsol}) at the center of a finite system with open boundary conditions and $J = 5000$ atoms. Fig.~\ref{unitcellsmoothplot}(a) and (b) show that $E_{\mathrm{sol}}$ approaches zero inversely proportionally with $\xi$~\cite{brzezicki20}.
The numerical data in Fig.~\ref{unitcellsmoothplot}(b), for $t/u_0 = 0.5$ and for $t/u_0 = 2.0$, coincides for $\xi \gg 1$, showing that $E_{\mathrm{sol}}/u_0$ is independent of $t/u_0$ in this regime.
We fit $\ln (|E_{\mathrm{sol}} / u_0|)$ versus $\ln \xi$ to a straight line $y = mx +c$ with data in the range $500 \leq \xi \leq 5000$. Repeating this for different data sets in the range $0.1 \leq t/u_0 \leq 10.0$ yields $m = -0.998 \pm 0.001$ and $c = -0.71 \pm 0.01$, so we deduce that
\begin{eqnarray}
E_{\mathrm{sol}} \approx - \frac{u_0}{2\xi} . \label{eunitsmooth}
\end{eqnarray}
This equation is shown as the solid line in Fig.~\ref{unitcellsmoothplot}(b).

Fig.~\ref{unitcellsmoothplot}(c) and (d) show polarization $p_y$ as a function of width $\xi$, and this numerical data is very similar to that of the atomically-smooth soliton, Fig.~\ref{cdw4plot}(c) and (d). For $\xi \gg 1$, the difference in the two sets of data is negligible. This demonstrates that $1 - p_y$ is inversely proportional to $\xi$, Eq.~(\ref{pysmooth}), and that $p_y$ doesn't depend on the microsopic profile of the soliton (atomically-smooth or unit-cell-smooth).

In the remainder of this paper, we present numerical data for atomically-smooth solitons~(\ref{smoothsol}) because the data for unit-cell-smooth solitons~(\ref{unitsmoothsol}) is qualitatively the same (for both soliton charge and robustness to disorder).
In Section~\ref{s:disorder}, we consider how robust the state localized on a soliton in the CDW phase is to different types of disorder. Before that, Section~\ref{s:charge}, we consider the electric charge of the soliton in the CDW phase.

\section{Electric charge of the soliton}\label{s:charge}

The electric charge of a soliton texture in $\Delta$ in the SSH phase is half-integer~\cite{jackiwrebbi76,heeger88} (for spinless electrons at half filling) and generally fractional~\cite{ricemele82,kivelson83,heeger88} for a soliton texture in $\Delta$ in the Rice-Mele model.
Now we determine the electric charge of the soliton in $u$ in the CDW phase. We consider a system of spinless electrons at half filling and at zero temperature, with electron charge $-e$ where $e>0$. For a finite system of $J$ atoms, the electronic charge is $-eJ/2$ in total.
The soliton level discussed in the previous sections is the highest occupied energy  level. Its eigenstate is normalized to unit probability over the whole sample so that, if one were to consider this state in isolation and integrate over the whole sample, one would incorrectly conclude that the soliton charge was $-e$. The presence of the soliton state disturbs the other states and, in determining soliton charge, it is necessary to sum over all valence band levels in the spatial vicinity of the soliton.

\begin{figure}[t]
\includegraphics[scale=0.38]{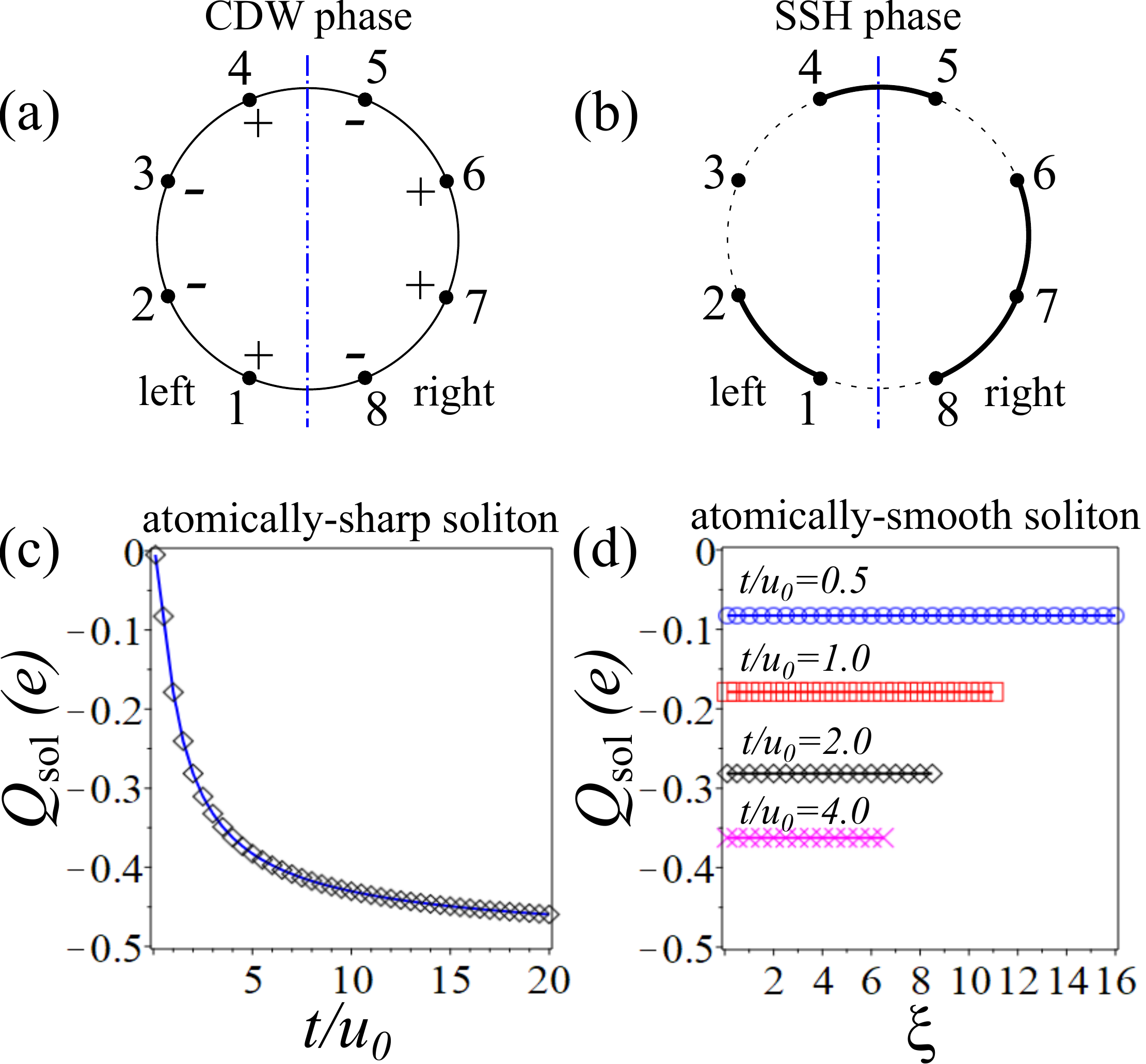}
\caption{(a), (b) Schematic of the set-up for determining soliton electric charge numerically~(\ref{chargenumeric}) with a soliton on the left and an antisoliton on the right of a system with periodic boundary conditions and $J=8$ atoms. (a) is for a texture in onsite energies in the CDW phase, indicated by plus and minus symbols, (b) is for a texture in hopping strengths in the SSH phase, indicated by solid and dashed lines. Dash-dot vertical blue lines show the boundary between the left and right sides.
(c), (d) Soliton electric charge $Q_{\mathrm{sol}}$ in the CDW phase determined for a system with $J = 5000$ atoms. (c) shows the charge of an atomically-sharp soliton as a function of $t/u_0$ with numerical data (diamonds) and the analytic formula~(\ref{solitoncharge})~\cite{brzezicki20} (solid line).
(d) Dependence of the charge on soliton width $\xi$ for an atomically-smooth soliton. In both plots, symbols are numerical data and solid lines are the analytic formula~(\ref{solitoncharge}). Data is only plotted until the point when the soliton energy $E_{\mathrm{sol}}$ is zero within numerical precision (see main text). Blue circles are $t/u_0 = 0.5$, red squares are $t/u_0 = 1.0$, black diamonds are $t/u_0 = 2.0$, and magenta crosses are $t/u_0 = 4.0$, 
}\label{charge1plot}
\end{figure}

In order to determine the soliton charge numerically, we consider a system in position space with periodic boundary conditions in order to eliminate any spurious end effects, Fig.~\ref{charge1plot}(a), (b). We then introduce a pair of a soliton and an antisoliton which are widely separated at the opposite sides of the ring, dubbed `left' and `right', respectively. Left ($L$) sites are $j = 1,2,\ldots,J/2$ and right ($R$) sites are $j = J/2 + 1,J/2 + 2,\ldots,J$. The antisoliton has the inverted texture of the soliton, i.e. it has the same magnitude of parameters $u$, $\xi$, etc, and is, therefore, assumed to have the opposite charge of the soliton. Then, the soliton charge is determined by calculating the difference in the total charges on the left and right sides of the ring by summing over all negative energy states with index $n = 1,2,\ldots,J/2$:
\begin{eqnarray}
Q_{\mathrm{sol}} = - \frac{e}{2} \sum_{n=1}^{J/2} \left[
\sum_{j \in L} | \psi_{n,j}|^2 - \sum_{j \in R} | \psi_{n,j}|^2 \right] . \label{chargenumeric}
\end{eqnarray}
The numerical procedure fails when the soliton and antisoliton energy levels are zero (within numerical precision) and, thus, degenerate because the corresponding eigenstates may be linear combinations of the two (and not solely localized on the left or the right). Such degeneracy can be broken by introducing an infinitesimal symmetry breaking e.g. a tiny value of $\Delta$ in the CDW phase. However, here we consider cases when the energy levels aren't degenerate (i.e. an atomically-sharp soliton or a smooth one with $\xi$ not too large).

The charge of an atomically-sharp soliton was determined analytically in Ref.~\cite{brzezicki20} by relating it to the charge of the ends of a chain in the pristine CDW phase (e.g. the charge of a soliton with two consecutive $-u_0$ onsite energies as in Fig.~\ref{dptplot}(a) can be related to the charge at the end of a chain that terminates with $-u_0$).
For completeness, we briefly outline this derivation in Appendix~\ref{a:charge}, which gives the charge of an atomically-sharp soliton~\cite{brzezicki20} as
\begin{eqnarray}
Q_{\mathrm{sol}} \approx - \frac{e}{2} \left[ 1 - \zeta ( u_0 ) \right] , \label{solitoncharge}
\end{eqnarray}
where 
\begin{eqnarray}
\zeta ( u_0 ) \approx \frac{2}{\pi} \frac{|u_0|}{\sqrt{u_0^2 + 4t^2}} K \bigg( \frac{2t}{\sqrt{u_0^2 + 4t^2}} \bigg) , \label{zetau0}
\end{eqnarray}
and $K(x)$ is the complete elliptic integral of the first kind,
\begin{eqnarray}
K(x) = \int_0^{\pi /2} \frac{d\theta}{\sqrt{1 - x^2 \sin^2 \theta}} .
\end{eqnarray}
The function $\zeta ( u_0 )$ describes the magnitude of the difference in probability densities $|\psi_B|^2 - |\psi_A|^2$ for the occupied valence bands, and it modifies $Q_{\mathrm{sol}}$ by describing an unequal distribution of charge between the two sublattices. Note that $\zeta (u_0) \geq 0$ by definition and it is independent of the sign of $u_0$. Function $K(x) = \pi / 2$ for $x \ll1$ and $K(x) \rightarrow \infty$ for $x \rightarrow 1$. This means that $\zeta ( u_0 ) \rightarrow 1$ for $u_0 \gg t$ and $\zeta ( u_0 ) \rightarrow 0$ for $u_0 \ll t$.
Hence $Q_{\mathrm{sol}} \rightarrow 0$ for $u_0 \gg t$ and $Q_{\mathrm{sol}} \rightarrow - e/2$ for $u_0 \ll t$ (for a sufficiently large system with $J \gg 1$).

Figure~\ref{charge1plot}(c),(d) show numerical data (symbols) for the soliton charge $Q_{\mathrm{sol}}$ for a system with $J = 5000$ atoms, and solid lines are fits to the analytic formula~(\ref{solitoncharge}). Figure~\ref{charge1plot}(c) is for an atomically-sharp soliton, showing dependence on the ratio $t/u_0$, and the agreement of numerics with analytics~(\ref{solitoncharge}) is extremely good.
As with $E_{\mathrm{sol}}$ and $p_y$ (discussed in Section~\ref{s:atomic}), it is possible to increase the ratio $t/u_0$ and approach the value $Q_{\mathrm{sol}} = -e/2$ expected for a topological system (as long as $t/u_0$ doesn't become huge $t/u_0 \sim J$). 

Numerically, we implement a smooth soliton and antisoliton pair, with the same width $\xi$, as
$u_j = (-1)^{j+1} u_0 + (-1)^{j} u_0 \tanh [ (j - j_1)/\xi ] + (-1)^{j+1} u_0 \tanh [ (j - j_2)/\xi ]$,
with centers $j_1$ and $j_2$.
Figure~\ref{charge1plot}(d) shows that $Q_{\mathrm{sol}}$ is independent of the soliton width $\xi$~\cite{smoothnote}, so that the analytic formula~(\ref{solitoncharge}) remains applicable even for a smooth soliton.
In this respect, $Q_{\mathrm{sol}}$ behaves quite differently to $E_{\mathrm{sol}}$ and $p_y$.
Note that we only plot numerical data until the point when the soliton energy becomes zero within numerical precision, because the numerical procedure fails when the soliton and antisoliton levels are degenerate.

The soliton charge is independent of the width and microscopic structure of the texture~\cite{smoothnote} in the CDW phase, and this can be understood by considering the regime of weak hopping ($t \ll u_0$). We discuss the example in Fig.~\ref{charge1plot}(a) with only $J=8$ atoms, a soliton centered between sites $2$ and $3$ on the left, and an antisoliton centered between sites $6$ and $7$ on the right. When $t =0$, the states $\psi_j$ are all localized on atoms with position index $j = 1,2,\ldots , J$, e.g. $\psi_2^T = \begin{pmatrix} 0 & 1 & 0 & 0 & 0 & 0  & 0 & 0 \end{pmatrix}$. Both left and right sides have two occupied valence band states $\psi_2$ and $\psi_3$, $\psi_5$ and $\psi_8$, so that the charge~(\ref{chargenumeric}) is zero.
For finite $t$ (with $t \ll u_0$), the valence band states associated with the soliton on the left are $\psi_{\mathrm{sol}}^{T} = \begin{pmatrix}
- \tau & 1 & 1 & - \tau & 0 & 0  & 0 & 0 \end{pmatrix} / \sqrt{2(1+\tau^2)}$ and 
$\psi_{\mathrm{sol}^{\prime}}^{T} = \begin{pmatrix}
\tau & 1 & - 1 & - \tau & 0 & 0  & 0 & 0 \end{pmatrix} / \sqrt{2(1+\tau^2)}$ with $\tau = t / (2u_0)$. These states are still fully localized on the left side, so still contribute a probability of two. However, the valence band states associated with the boundary sites $5$ and $8$ extend across the boundary as
$\psi_{5}^{T} = \begin{pmatrix}
0 & 0 & 0 & - \tau & 1 & - \tau  & 0 & 0 \end{pmatrix} / \sqrt{1+2\tau^2}$ and 
$\psi_{8}^{T} = \begin{pmatrix}
-\tau & 0 & 0 & 0 & 0 & 0- \tau & 1 \end{pmatrix} / \sqrt{1+2\tau^2}$.
Thus, the presence of the soliton and antisoliton modifies other valence band states, including those at the boundary, which leads to a motion of charge between the right and left sides. Using Eq.~(\ref{chargenumeric}), $Q_{\mathrm{sol}} = - 2e \tau^2 = -e t^2 / (2 u_0^2)$ which agrees with the atomically-sharp result~(\ref{solitoncharge}) to lowest order in $t/u_0$. Note that this estimate depends only on the magnitude of the texture $u_0$ at the boundary as it appears in $\psi_{5}$ and $\psi_{8}$, and is independent of the microscopic details of the texture near the soliton center, including the texture shape and width $\xi$.

We can consider soliton charge in the SSH phase in a similar way, using the fully-dimerized limit $\Delta = 2t$~\cite{asboth16}.
Fig.~\ref{charge1plot}(b) shows a soliton centered on site $3$ on the left, and an antisoliton centered on site $7$ on the right, solid lines correspond to non-zero hopping (magnitude $2t$), dashed lines to zero hopping (fully broken bonds). 
The isolated state on site $3$ is at $E=0$, the dimers give a valence band state at $E = - 2t$ with probability equally distributed on both sites, and the trimer (sites $6$, $7$, $8$) has a valence band state at $E = - 2t$ and a state at $E=0$ (plus a conduction band state). For the valence band states at $E = - 2t$, both sides have a contribution to the probability of $3/2$ (half integer because of the dimer state on sites $4$ and $5$). There are also two $E=0$ states: one on the soliton on the left and one on the antisoliton on the right, but only one of them may be occupied. Thus, the charge of the soliton on the left is $Q_{\mathrm{sol}} = \pm e / 2$~\cite{jackiwrebbi76,kivelson83,heeger88} where the plus (minus) sign is for when it is unoccupied (occupied).

\begin{table}[t]
\begin{center}
\caption{\label{tablecharge}Soliton electric charge in the CDW phase depending on the particular soliton state where $Q$ is charge in a spinless system (rows $2$-$5$), $\tilde{Q}$ is charge in a spinful system (rows $6$-$11$). Subscript `sol' (`antisol') indicates a soliton (antisoliton) and superscript $0$, $1$, or $2$ indicates the occupancy of the soliton state. The function $\zeta (u_0)$ is an analytical approximation~\cite{brzezicki20} given in Eq.~(\ref{zetau0}). The electron charge is $-e$ where $e>0$.}
\begin{tabular}{ L{1.2cm} | C{2.0cm} | C{1.0cm} | C{1.0cm} }
\hline
state & charge & $u_0 \gg t$ & $u_0 \ll t$ \\
\hline \hline
$Q_{\mathrm{sol}}^{(0)}$ & $e (1+\zeta) / 2$ & $e$ & $e/2$ \\[2pt]
\hline
$Q_{\mathrm{sol}}^{(1)}$ & $- e (1-\zeta) / 2$ & $0$ & $-e/2$ \\[2pt]
\hline
$Q_{\mathrm{antisol}}^{(0)}$ & $e (1-\zeta) / 2$ & $0$ & $e/2$ \\[2pt]
\hline
$Q_{\mathrm{antisol}}^{(1)}$ & $-e (1+\zeta) / 2$ & $-e$ & $-e/2$ \\[2pt]
\hline\hline
$\tilde{Q}_{\mathrm{sol}}^{(0)}$ & $e (1+\zeta)$ & $2e$ & $e$ \\[2pt]
\hline
$\tilde{Q}_{\mathrm{sol}}^{(1)}$ & $e \zeta$ & $e$ & $0$ \\[2pt]
\hline
$\tilde{Q}_{\mathrm{sol}}^{(2)}$ & $- e (1-\zeta)$ & $0$ & $-e$ \\[2pt]
\hline
$\tilde{Q}_{\mathrm{antisol}}^{(0)}$ & $e (1-\zeta)$ & $0$ & $e$ \\[2pt]
\hline
$\tilde{Q}_{\mathrm{antisol}}^{(1)}$ & $-e \zeta$ & $-e$ & $0$ \\[2pt]
\hline
$\tilde{Q}_{\mathrm{antisol}}^{(2)}$ & $- e (1+\zeta)$ & $-2e$ & $-e$ \\[2pt]
\hline
\end{tabular}
\end{center}
\end{table}

Here we have considered spinless electrons at half filling. For spinless electrons in the CDW phase, the charge $Q^{(n)}$ for different occupancy of the soliton state ($n=0,1$) may be found from $Q_{\mathrm{sol}}^{(1)}$, Eq.~(\ref{solitoncharge}), and $Q_{\mathrm{antisol}}^{(0)} = - Q_{\mathrm{sol}}^{(1)}$ by adding or subtracting electric charge $\mp e$~\cite{brzezicki20}, and the results are summarized in rows $2$-$5$ of Table~\ref{tablecharge}.
For spinful electrons, we denote soliton charge as $\tilde{Q}$, and twofold spin degeneracy gives 
$\tilde{Q}_{\mathrm{sol}}^{(2)} = 2 Q_{\mathrm{sol}}^{(1)}$
and
$\tilde{Q}_{\mathrm{antisol}}^{(0)} = 2 Q_{\mathrm{antisol}}^{(0)}$.
Then, by adding or subtracting electric charge $\mp e$ we find the soliton charge for different occupancy ($n=0,1,2$), and the results are summarized in rows $6$-$11$ of Table~\ref{tablecharge}.
We introduce parameter $n_0$ such that $n_0+1$ is equal to the number of different possible occupations of the state, i.e. $0 \leq n \leq n_0$.
Then, the results for the spinless case ($n_0 = 1$) and the spinful case ($n_0 = 2$) may be combined as $Q_{\mathrm{sol}}^{(n)} \approx e [n_0 (1+\zeta ) - 2n]/2$ and $Q_{\mathrm{antisol}}^{(n)} \approx e [n_0 (1-\zeta ) - 2n]/2$.
In the large bandwidth limit $u_0 \ll t$ (fourth column of Table~\ref{tablecharge}), then $\zeta \rightarrow 0$ and the soliton and antisoliton charges with the same occupancy are equal, and they coincide with the known values for topological solitons in the SSH phase~\cite{kivelson83} as described by $Q^{(n)} = e (n_0 - 2n)/2$.

\section{Disorder and sample-to-sample parameter variations}\label{s:disorder}

\subsection{Nonsymmorphic chiral symmetry in position space}

We begin by considering the general form of a $J \times J$ Hamiltonian $H$ in position space which satisfies nonsymmorphic chiral symmetry $S_y^{-1} H S_y = - H$. For even $J$, it may be written generically as
\begin{eqnarray}
H = \begin{pmatrix}
h_1 & h_2 & h_3 & \hdots & h_3^{\ast} & h_2^{\ast} \\
h_2^{\ast} & -h_1 & h_2 & \hdots & h_4^{\ast} & -h_3^{\ast} \\
h_3^{\ast} & h_2^{\ast} & h_1 & \hdots & h_5^{\ast} & h_4^{\ast} \\
\vdots & \vdots & \vdots & \vdots & \vdots & \vdots \\
h_3 & h_4 & h_5 & \hdots & h_1 & h_2 \\
h_2 & -h_3 & h_4 & \hdots & h_2^{\ast} & -h_1
\end{pmatrix} , \label{hns}
\end{eqnarray}
where $h_1, h_2, \ldots , h_{J/2 +1}$ are arbitrary components. With the property of hermicity, there are $J$ real numbers: $h_1$ and $h_{J/2 +1}$ are real, and the other components, $h_2, h_3, \ldots , h_{J/2}$, are complex. The CDW phase satisfies this symmetry, but only with periodic boundary conditions, and no textures in the components. Note that it is not possible to write a non-zero Hamiltonian that satisfies this symmetry for odd $J$.

To satisfy the nonsymmorphic chiral symmetry, the components $h_i$ must be uniform across the entire sample. Thus it is not possible to have microscopic disorder within a sample; in order to satisfy the symmetry, variations must be restricted to parameter values that differ from sample to sample within an ensemble (e.g. as in gate-induced variations). In the following, we consider the robustness of the soliton states in a finite system to four different types of disorder~\cite{inui94,perezgonzalez19,scollon20} or sample-to-sample variatons: (i) `onsite disorder' (diagonal) gives an additional contribution to the onsite energy of site $j = 1,2,\ldots,J$ as $\delta \epsilon_j$, where $\delta \epsilon_j$ is drawn randomly from a uniform distribution $-W \leq \delta \epsilon_j \leq W$ with disorder strength $W$; (ii) `hopping disorder' (off-diagonal) gives an additional contribution to the nearest-neighbor hopping between site $j$ and $j+1$ of $\delta t_j$, $j = 1,2,\ldots,J-1$, where $\delta t_j$ is drawn randomly from a uniform distribution $-W \leq \delta t_j \leq W$. For a given $j$, both $\delta \epsilon_j$ and $\delta t_j$ also vary between members of the ensemble. (iii) `onsite variations' give an additional contribution to the staggered onsite energy $\delta u$ that is uniform across the entire sample, but is drawn randomly from a uniform distribution $-W \leq \delta u \leq W$ for different ensemble members; (iv) `hopping variations' give an additional contribution to the nearest-neighbor hopping $\delta t$ that is uniform across the entire sample, but is drawn randomly from a uniform distribution $-W \leq \delta t \leq W$ for different ensemble members.

\begin{figure}[t]
\includegraphics[scale=0.40]{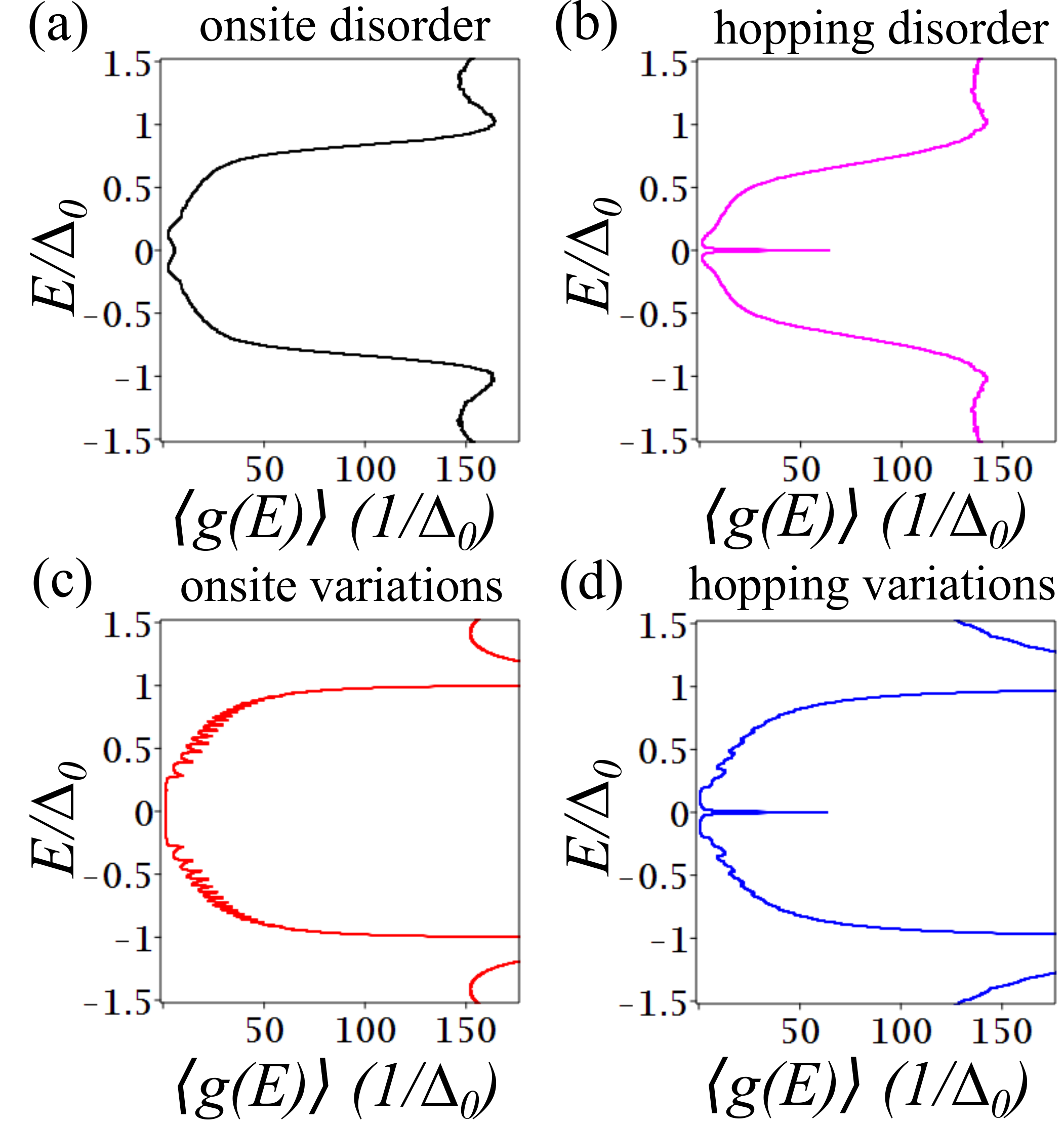}
\caption{Dependence of the disorder-averaged density of states $\langle g(E) \rangle$ on energy $E$ in the SSH phase for a single soliton~(\ref{sshsol}) with width $\xi = 50$, $t/\Delta_0 = 1$, at the center of a finite system with open boundary conditions and $J = 501$ atoms. For all plots, the disorder strength is $W/\Delta_0 = 0.5$, and $\langle g(E) \rangle$ is determined using Eq.~(\ref{dos}) with broadening $\delta = 0.005\Delta_0$. (a) is for onsite disorder (black), (b) is hopping disorder (magenta), (c) is onsite variations (red), and (d) is hopping variations (blue). Averaging is done with respect to $10,000$ disorder realizations.
}\label{dos2plot}
\end{figure}

\begin{figure}[t]
\includegraphics[scale=0.21]{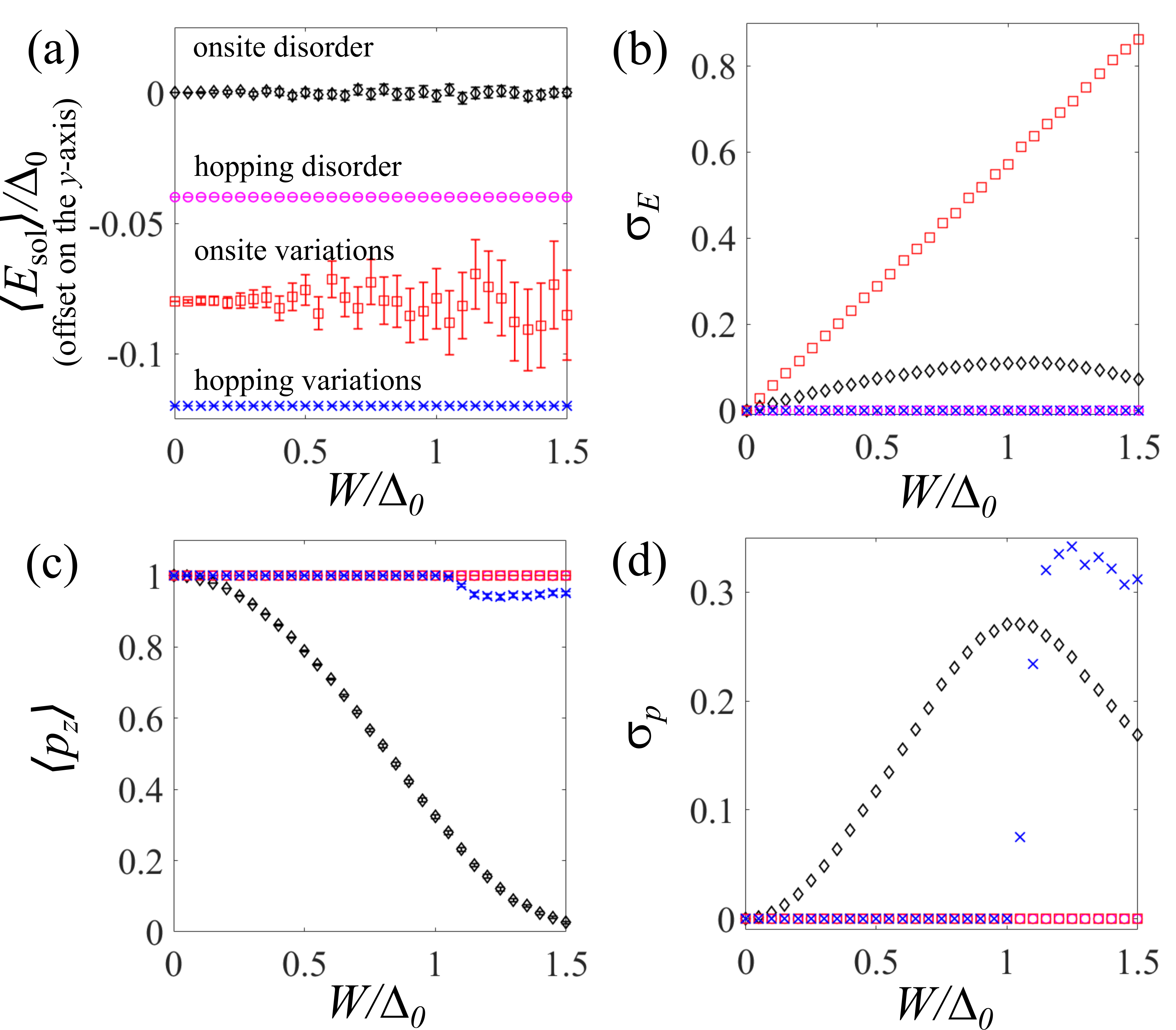}
\caption{Dependence in the SSH phase on disorder strength $W$ for a single soliton~(\ref{sshsol}) with width $\xi = 50$, $t/\Delta_0 = 1$, at the center of a finite system with open boundary conditions and $J = 501$ atoms. In all plots, black diamonds show numerical data for onsite disorder, magenta circles show hopping disorder, red squares show onsite variations, and blue crosses show hopping variations. Averaging is done with respect to $10,000$ disorder realizations. (a) shows the mean soliton energy $\langle E_{\mathrm{sol}} \rangle$ with error bars, with each data set offset from zero by multiples of $0.04\Delta_0$. (b) shows the standard deviation $\sigma_E$ of $E_{\mathrm{sol}}$. (c) shows the mean polarization $\langle p_z \rangle$ with error bars. (d) shows the standard deviation $\sigma_p$ of $p_z$.
}\label{dis2plot}
\end{figure}

As an example, onsite disorder in the CDW phase (without solitons) would give onsite energies $u + \delta \epsilon_1 , - u + \delta \epsilon_2 , + u + \delta \epsilon_3 ,  \ldots , - u + \delta \epsilon_J$ for the first member of the ensemble, $u + \delta \epsilon_1^{\prime} , - u + \delta \epsilon_2^{\prime} , + u + \delta \epsilon_3^{\prime} ,  \ldots , - u + \delta \epsilon_J^{\prime}$ for the second, $u + \delta \epsilon_1^{\prime\prime} , - u + \delta \epsilon_2^{\prime\prime} , + u + \delta \epsilon_3^{\prime\prime} ,  \ldots , - u + \delta \epsilon_J^{\prime\prime}$ for the third, etc., where $\delta \epsilon_j \neq \delta \epsilon_j^{\prime} \neq \delta \epsilon_j^{\prime\prime}$.
However, onsite variations in the same phase would give onsite energies $(u + \delta u) , - (u + \delta u) , + (u + \delta u) ,  \ldots , - (u + \delta u)$ for the first member of the ensemble, $(u + \delta u^{\prime}) , - (u + \delta u^{\prime}) , + (u + \delta u^{\prime}) ,  \ldots , - (u + \delta u^{\prime})$ for the second, $(u + \delta u^{\prime\prime}) , - (u + \delta u^{\prime\prime}) , + (u + \delta u^{\prime\prime}) ,  \ldots , - (u + \delta u^{\prime\prime})$ for the third, etc., where $\delta u \neq \delta u^{\prime} \neq \delta u^{\prime\prime}$.

\begin{figure}[t]
\includegraphics[scale=0.40]{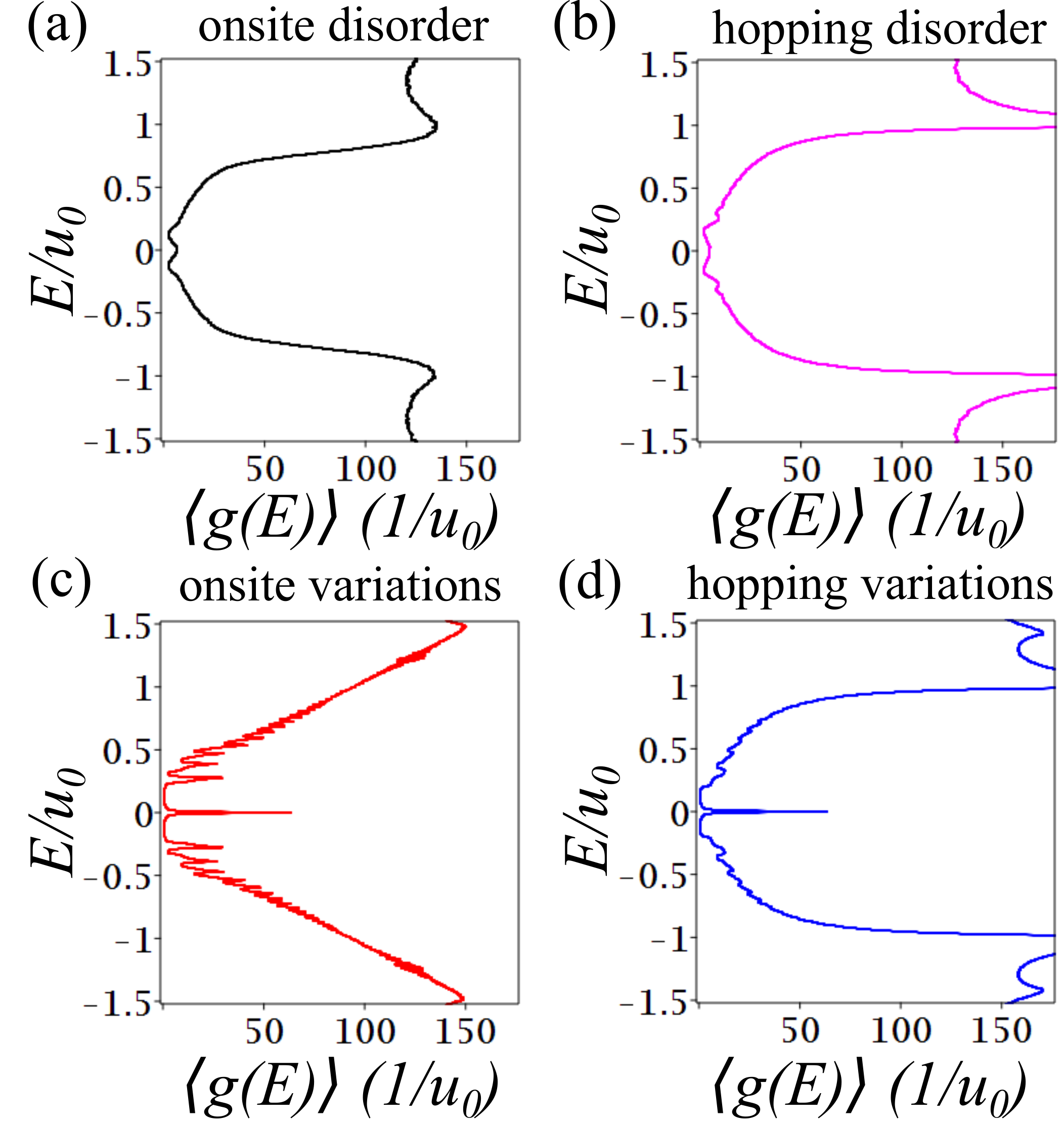}
\caption{Dependence of the disorder-averaged density of states $\langle g(E) \rangle$ on energy $E$ in the CDW phase for a single soliton~(\ref{smoothsol}) with width $\xi = 50$, $t/u_0 = 1$, at the center of a finite system with open boundary conditions and $J = 500$ atoms. For all plots, the disorder strength is $W/u_0 = 0.5$, and $\langle g(E) \rangle$ is determined using Eq.~(\ref{dos}) with broadening $\delta = 0.005u_0$. (a) is for onsite disorder (black), (b) is hopping disorder (magenta), (c) is onsite variations (red), and (d) is hopping variations (blue). Averaging is done with respect to $10,000$ disorder realizations.
}\label{dos1plot}
\end{figure}

\begin{figure}[t]
\includegraphics[scale=0.21]{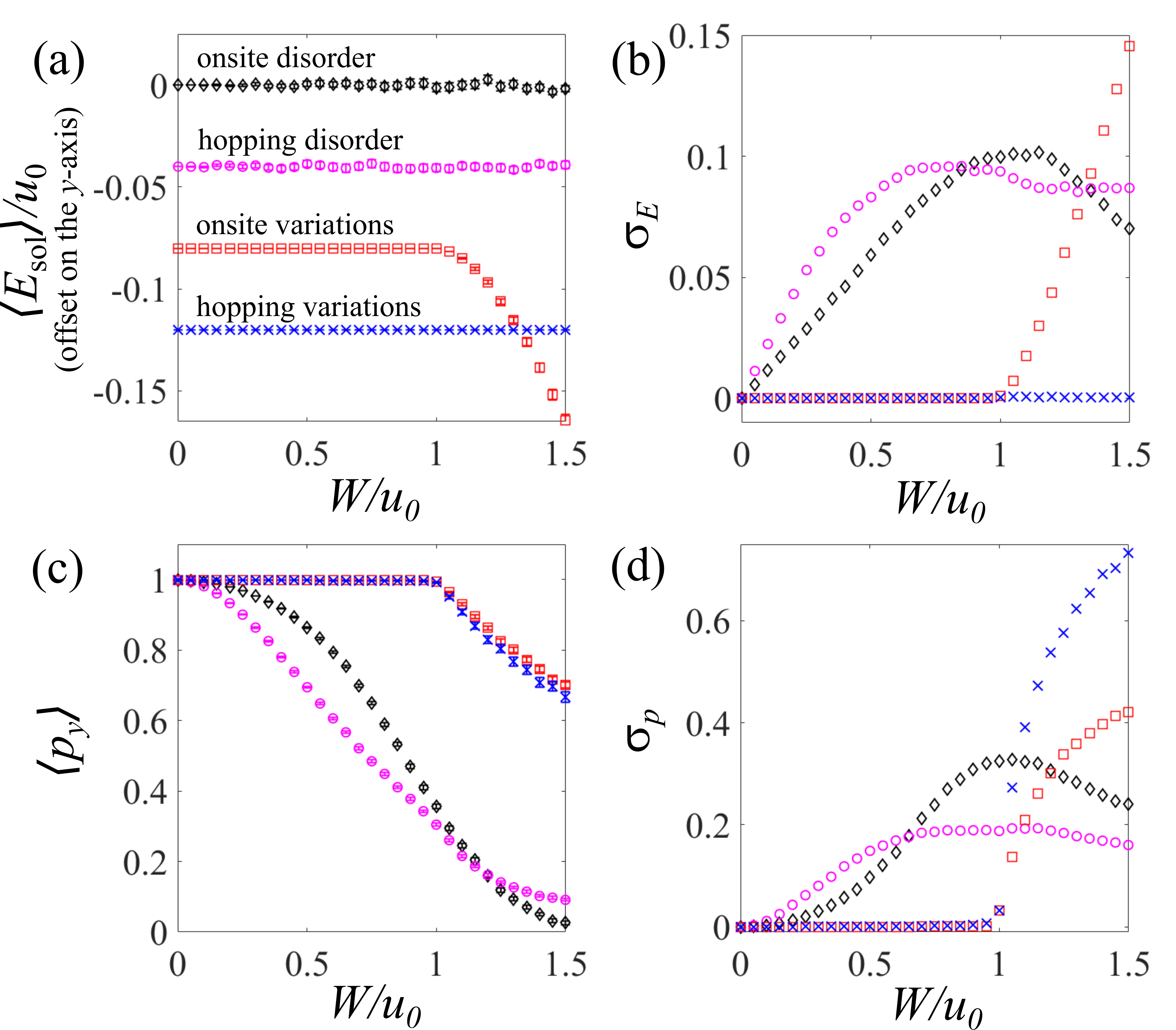}
\caption{Dependence in the CDW phase on disorder strength $W$ for a single soliton~(\ref{smoothsol}) with width $\xi = 50$, $t/u_0 = 1$, at the center of a finite system with open boundary conditions and $J = 500$ atoms. In all plots, black diamonds show numerical data for onsite disorder, magenta circles show hopping disorder, red squares show onsite variations, and blue crosses show hopping variations. Averaging is done with respect to $10,000$ disorder realizations. (a) shows the mean soliton energy $\langle E_{\mathrm{sol}} \rangle$ with error bars, with each data set offset from zero by multiples of $0.04u_0$. (b) shows the standard deviation $\sigma_E$ of $E_{\mathrm{sol}}$. (c) shows the mean polarization $\langle p_y \rangle$ with error bars. (d) shows the standard deviation $\sigma_p$ of $p_y$.
}\label{dis1plot}
\end{figure}

\subsection{Numerical methodology}

For a given disorder realization, the Hamiltonian is diagonalized, and the states are ordered from lowest energy upwards with labels $n = 1,2,\ldots , J$. Averages with respect to disorder are made with an ensemble of $10,000$ disorder realizations, and the properties (e.g. energy and polarization) of the levels with the same label $n$ are averaged. In the results we present, we focus on the soliton state with label $n = J/2$ for even $J$, or $n = (J+1)/2$ for odd $J$.
We present results for the mean soliton energy $\langle E_{\mathrm{sol}} \rangle$ and its standard deviation $\sigma_E$, and the mean polarization $\langle p_y \rangle$ and its standard deviation $\sigma_p$.

\subsection{Solitons in the SSH phase}

Before considering a soliton in the CDW phase, we consider an exemplar of preservation of bulk topology~\cite{perezgonzalez19,scollon20,vanmiert20}, namely a single soliton in the SSH phase ($u_0 = 0$) for a system with an odd number of atoms and stronger bonds at the ends. It supports a single localized state at zero energy~\cite{inui94}.
For a nearest-neighbor bond $t_{\ell}$ with index $\ell = 1 , 2, \ldots , J-1$,
\begin{eqnarray}
\Delta_{\ell} = (-1)^{\ell} \Delta_0 \tanh \left( \frac{\ell - \ell_0}{\xi} \right) , \label{sshsol}
\end{eqnarray}
where $\Delta_0$ is the magnitude at infinity, and $\xi$ is the width in dimensionless units written as the physical width divided by the atomic spacing ($a/2$). We consider the soliton to be centred on an atomic site, with integer site index $j_0$, such that the bond index is $\ell_0 = j_0 - 1/2$.

Figure~\ref{dos2plot} shows the mean density of states for a single soliton at the center of a SSH system with $J=501$ atoms, width $\xi = 50$, and disorder strength $W / \Delta_0 = 0.5$ for the four types of disorder and sample-to-sample variations.
Note that an integral of the density of states over a small energy window centered on each peak at $E=0$ will yield unity, reflecting the fact that there is a single soliton state.
For hopping disorder and hopping variations, the soliton level at $E=0$ is clearly visible, which is to be expected because hopping disorder conserves the SSH chiral symmetry. 
The level is barely visible for onsite disorder and is not discernible for onsite variations; staggered onsite energies break the SSH chiral symmetry.

\begin{figure}[t]
\includegraphics[scale=0.40]{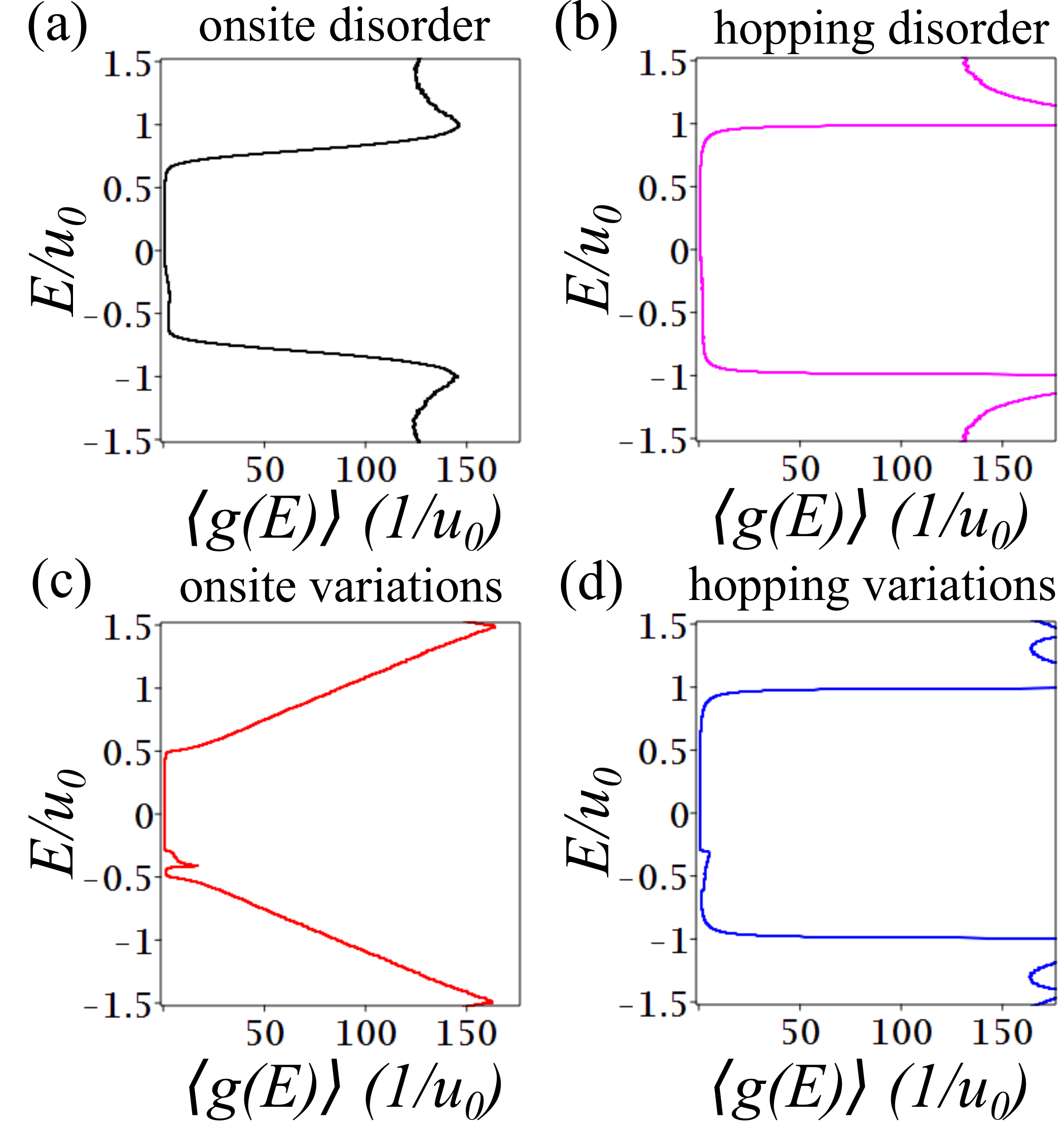}
\caption{Dependence of the disorder-averaged density of states $\langle g(E) \rangle$ on energy $E$ in the CDW phase for a single atomically-sharp soliton with $t/u_0 = 1$, at the center of a finite system with open boundary conditions and $J = 500$ atoms. For all plots, the disorder strength is $W/u_0 = 0.5$, and $\langle g(E) \rangle$ is determined using Eq.~(\ref{dos}) with broadening $\delta = 0.005u_0$. (a) is for onsite disorder (black), (b) is hopping disorder (magenta), (c) is onsite variations (red), and (d) is hopping variations (blue). Averaging is done with respect to $10,000$ disorder realizations.
}\label{dos3plot}
\end{figure}

\begin{figure}[t]
\includegraphics[scale=0.21]{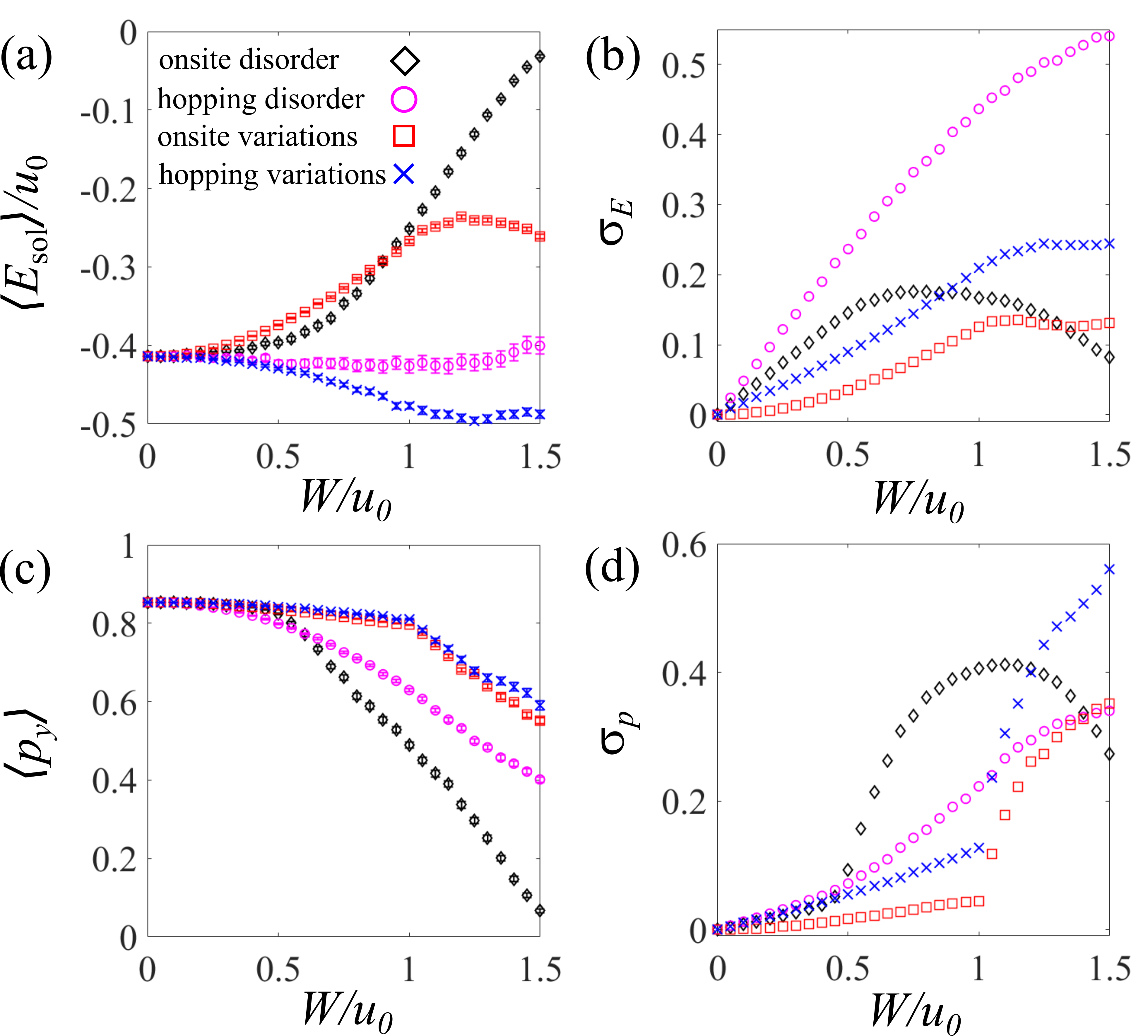}
\caption{Dependence in the CDW phase on disorder strength $W$ for a single atomically-sharp soliton, $t/u_0 = 1$, at the center of a finite system with open boundary conditions and $J = 500$ atoms. In all plots, black diamonds show numerical data for onsite disorder, magenta circles show hopping disorder, red squares show onsite variations, and blue crosses show hopping variations. Averaging is done with respect to $10,000$ disorder realizations. (a) shows the mean soliton energy $\langle E_{\mathrm{sol}} \rangle$ with error bars. (b) shows the standard deviation $\sigma_E$ of $E_{\mathrm{sol}}$. (c) shows the mean polarization $\langle p_y \rangle$ with error bars. (d) shows the standard deviation $\sigma_p$ of $p_y$.
}\label{dis3plot}
\end{figure}

Figure~\ref{dis2plot} shows properties of the SSH soliton state as a function of disorder strength.
For hopping disorder, mean energy $\langle E_{\mathrm{sol}} \rangle = 0$, polarization $\langle p_z \rangle = 1$, and standard deviations are zero (within numerical precision) for the disorder strengths we consider (up to $W/\Delta_0 = 1.5$). For hopping variations, $\langle E_{\mathrm{sol}} \rangle = 0$ with zero standard deviation, but $\langle p_z \rangle = 1$ only up to $W/\Delta_0 = 1$. After this, $\langle p_z \rangle$ is slightly less than one, and the standard deviation $\sigma_p$ is non-zero.
The system actually remains topological, but some members of the ensemble have bonds at their ends that are weaker (in magnitude) than the adjacent bonds. They thus support additional zero energy states at the ends: the two new states have $p_z = 1$ while the polarization of the central soliton flips to $p_z =-1$. The addition of contributions with $p_z = -1$ reduces the ensemble average $\langle p_z \rangle$. Note also that $\langle p_z \rangle = 1$ ($\sigma_p = 0$) for onsite variations even though $\sigma_E \neq 0$.

\subsection{Solitons in the CDW phase}

We now compare solitons in the CDW phase to those in the SSH phase.
Figure~\ref{dos1plot} shows the mean density of states for a single soliton at the center of a CDW system with $J=500$ atoms, width $\xi = 50$, and disorder strength $W / \Delta_0 = 0.5$ for the four types of disorder and sample-to-sample variations~\cite{smoothnote}.
For the sample-to-sample variations (onsite and hopping), the soliton level at $E=0$ is clearly visible, but it is barely visible for spatial disorder (onsite and hopping).
Figure~\ref{dis1plot} shows properties of the CDW soliton state as a function of disorder strength~\cite{smoothnote}. For zero disorder, this state is at zero energy (within numerical precision) with $p_y = 0.9975$. It is clearly fragile in the presence of spatial disorder (onsite and hopping) with non-zero standard deviations $\sigma_E$ and $\sigma_p$, and $\langle p_y \rangle$ approaching zero for large disorder. However, for sample-to-sample variations (onsite and hopping), the level remains at zero energy with high $p_y$ values and $\sigma_E \approx \sigma_p \approx 0$ until $W/u_0 \approx 1$.
For $W/u_0 \agt 1$ and onsite variations, some of the ensemble members have $|\delta u| \agt u_0$ which destroys the state localized on the soliton. 
For $W/u_0 \agt 1$ and hopping variations, some of the ensemble members have total hopping that is negative. For these, the soliton energy is still near zero, but there is a flip in the sign of $p_y$ (to $p_y \approx -1$), resulting in a decrease of $\langle p_y \rangle$ and an increase in $\sigma_p$.

The mean density of states for an atomically-sharp single soliton at the center of a CDW system with $J=500$ atoms and disorder strength $W / \Delta_0 = 0.5$ is shown in Figure~\ref{dos3plot}. The soliton state is barely visible for spatial disorder (onsite and hopping). It can be seen for sample-to-sample variations (onsite and hopping), but with significant width (and an asymmetric shape as a function of $E$). 
Figure~\ref{dis3plot} shows properties of the atomically-sharp CDW soliton state as a function of disorder strength. For zero disorder, this state is at non-zero energy $E_{\mathrm{sol}} = -0.4142 u_0$ with $p_y = 0.8536$. To some extent the behavior mirrors that of a smooth soliton, Fig.~\ref{dis1plot}, but it is less robust.
For sample-to-sample variations (onsite and hopping), $\sigma_E$ and $\sigma_p$ are smaller for $W/u_0 \alt 1$ than for disorder, and $\langle p_y \rangle$ remains close to its original value for $W/u_0 \alt 1$.
For onsite disorder and variations, the mean energy $\langle E_{\mathrm{sol}} \rangle$ actually moves towards zero as disorder increases, which we attribute to a general narrowing of the mean band gap, as seen in Figure~\ref{dos3plot} (by comparing the onsite with the hopping figures).

\section{Conclusion}

In the charge density wave (CDW) phase with staggered onsite hopping, chiral symmetry is nonsymmorphic so that an end, spatial disorder, or a spatial texture in parameter values break the chiral symmetry. Despite this, an atomically-sharp soliton supports a localized state with an energy $E_{\mathrm{sol}}$ which lies within the band gap for a wide range of parameter values.
Increasing the ratio of the bandwidth to band gap (i.e. the ratio $t/u_0$), in a sufficiently long system, can drive the soliton energy $E_{\mathrm{sol}}$ towards zero, the polarization $p_y$ towards one, and the occupied soliton electric charge $Q_{\mathrm{sol}}$ towards $-e/2$.

For a smooth soliton of width $\xi$, the dependence of the energy level $E_{\mathrm{sol}}$ on $\xi$ depends on microscopic details of the soliton texture~\cite{brzezicki20}: for an atomically-smooth soliton~(\ref{smoothsol}), $E_{\mathrm{sol}}$ scales to zero exponentially with $\xi$ whereas, for a unit-cell-smooth soliton~(\ref{unitsmoothsol}), $E_{\mathrm{sol}}$ is inversely proportional to $\xi$. However, both types of smooth soliton share the same dependences of their polarization $p_y$ and charge $Q_{\mathrm{sol}}$ on $\xi$: 
$p_y$ approaches one only inversely proportionally with $\xi$, and $Q_{\mathrm{sol}}$ is independent of $\xi$.
Hence, any soliton in a finite system with open boundary conditions cannot be regarded as topological. Nevertheless, a smooth soliton in the CDW phase can be robust with respect to sample-to-sample variations in the staggered onsite energies and nearest-neighbor hoppings.

All relevant data present in this publication can be accessed at \cite{datalink}.

\begin{acknowledgments}
The authors thank A. Romito and H. Schomerus for helpful discussions.
\end{acknowledgments}

\appendix

\section{The energy level of a single atomically-sharp soliton in the CDW phase is at the band edge for $t/u_0=J/2$}\label{a:bandedge}

We consider a single atomically-sharp soliton in the CDW phase, placed at the centre of a system with $N$ unit cells (where $N$ is even), $J = 2N$ atoms, and open boundary conditions.
The aim is to demonstrate that there is an energy level exactly at the band edge $E = - u_0$ for $t/u_0 = J/2$.

The energy eigenvalue equation $H \psi = E \psi$, where $\psi$ is a $J$-component column vector of atomic states $\psi_j$, $j = 1,2,\ldots, J$, yields $J$ simultaneous equations. With $E = - u_0$, half of the equations give relations between pairs of components which may be summarized as
\begin{eqnarray}
\psi_1 = - \psi_3 = \psi_5 = \ldots &=& (-1)^{J/4} \psi_{J/2 + 1} , \label{psi1} \\
\psi_J = - \psi_{J-2} = \psi_{J-4} = \ldots &=& (-1)^{J/4} \psi_{J/2} . \label{psij}
\end{eqnarray}
The other $J/2$ simultaneous equations split into relations between the sites before ($j \leq J/2$) and after ($j > J/2$) the soliton. For example, the former are
\begin{eqnarray*}
t \psi_2 &=& - 2u_0 \psi_1 , \\
t \psi_2 + t \psi_4 &=& - 2u_0 \psi_3 , \\
t \psi_4 + t \psi_6 &=& - 2u_0 \psi_5 , \\
&\vdots& \\
t \psi_{J/2-2} + t \psi_{J/2} &=& - 2u_0 \psi_{J/2-1} .
\end{eqnarray*}
Using Eqs.~(\ref{psi1},\ref{psij}), these may be written as
\begin{eqnarray*}
- t \psi_2 &=& 2u_0 \psi_1 , \\
t \psi_2 + t \psi_4 &=& 2u_0 \psi_1 , \\
- t \psi_4 - t \psi_6 &=& 2u_0 \psi_1 , \\
&\vdots& \\
(-1)^{J/4} t \psi_{J/2-2} + t \psi_{J} &=& 2u_0 \psi_{1} .
\end{eqnarray*}
These $J/4$ equations may be added together to give
\begin{eqnarray}
t \psi_{J} = (Ju_0/2) \psi_1 .
\end{eqnarray}
Likewise, the $J/4$ simultaneous equations which give relations between the sites after ($j > J/2$) the soliton yield
\begin{eqnarray}
t \psi_{1} = (Ju_0/2) \psi_J .
\end{eqnarray}
These latter two equations are only compatible if $t/u_0 = J/2$.

\section{Analytic expression for the charge of an atomically-sharp soliton in the CDW phase}\label{a:charge}

We briefly outline the derivation of Ref.~\cite{brzezicki20} for the expression for the charge of an atomically-sharp soliton, Eq.~(\ref{solitoncharge}).
Consider the position space Hamiltonian of a pristine CDW chain with $j = 1,2,\ldots , J$ atoms and open boundary conditions as in Eq.~(\ref{rm2}) with $\Delta = 0$ and where the first site has onsite energy $u = u_0$.
This Hamiltonian may be block diagonalized using the eigenstates of a monoatomic chain (with $u_0=0$) which has eigenvalues and eigenstates as
\begin{eqnarray}
E_{n,s}^{(0)} &=& s \epsilon_n^{(0)} , \quad \epsilon_n^{(0)} = 2t \cos \! \left( \frac{n \pi}{J+1} \right) , \\
\psi_{n,s,j}^{(0)} &=& s^{j+1} \sqrt{\frac{2}{J+1}} \sin \! \left( \frac{n \pi j}{J+1} \right) ,
\end{eqnarray}
where $n = 1,2,\ldots J/2$ and $s = \pm 1$. In a basis with pairs $\psi_{n,+,j}^{(0)}$, $\psi_{n,-,j}^{(0)}$ for each $n$, the CDW Hamiltonian is reduced to $2 \times 2$ blocks,
\begin{eqnarray}
H_{n}^{2 \times 2} = \begin{pmatrix}
\epsilon_n^{(0)} & u_0 \\
u_0 & -\epsilon_n^{(0)}
\end{pmatrix} .
\end{eqnarray}
These have eigenvalues
\begin{eqnarray}
E_{n,\pm} = \pm \epsilon_n , \quad \epsilon_n =
\sqrt{u_0^2 + 4t^2 \cos^2 \!\left( \frac{n \pi}{J+1} \right)} .
\end{eqnarray}
For negative energy states, probability densities are
\begin{eqnarray}
|\psi_{n,-,j}|^2 = \frac{2}{J+1} \sin^2 \! \left( \frac{n \pi j}{J+1} \right)
\left[ 1 - (-1)^{j+1} \frac{u_0}{\epsilon_n} \right] . \label{valend}
\end{eqnarray}

The charge $Q_{\mathrm{end}}$ at the left end of the chain is determined by summing over all negative energy states and over $M$ sites near the end (where $M$ is an even number and $M \gg 1$):
\begin{eqnarray}
Q_{\mathrm{end}} = q_{\mathrm{e}} \sum_{n=1}^{J/2} \sum_{j=1}^{M} |\psi_{n,-,j}|^2 - q_{\mathrm{e}} \frac{M}{2} ,
\end{eqnarray}
where the electron charge is $q_{\mathrm{e}} = -e$, $e>0$.
The second term indicates that the charge is measured with respect to the contribution of $M$ sites with a perfectly homogeneous charge distribution (for half filling).
With the form of the probabilities~(\ref{valend}), and $M \gg 1$,
\begin{eqnarray}
Q_{\mathrm{end}} \approx  - \mathrm{sign} (u_0)\frac{q_{\mathrm{e}}}{4} \left[ 1 - \zeta ( u_0 ) \right] ,
\end{eqnarray}
where~\cite{brzezicki20}
\begin{eqnarray}
\zeta ( u_0 ) &=& \frac{2}{J+1} \sum_{n=1}^{J/2} \frac{|u_0|}{\epsilon_n} \\
&=& \frac{2}{J+1} \sum_{n=1}^{J/2} \frac{|u_0|}{\sqrt{u_0^2 + 4t^2 \cos^2 \! \left( \frac{n \pi}{J+1} \right)}} .
\end{eqnarray}
For a system with a large number of sites $J \gg 1$, this may be approximated as an integral,
\begin{eqnarray}
\zeta ( u_0 ) \approx \frac{a |u_0|}{\pi} \int_0^{\pi} \frac{dk}{\sqrt{u_0^2 + 4t^2 \cos^2 \! \left( \frac{k a }{2} \right)}} .
\end{eqnarray}
The complete elliptic integral of the first kind $K(x)$ is defined as
\begin{eqnarray}
K(x) = \int_0^{\pi /2} \frac{d\theta}{\sqrt{1 - x^2 \sin^2 \! \theta}} ,
\end{eqnarray}
so that
\begin{eqnarray}
\zeta ( u_0 ) \approx \frac{2}{\pi} \frac{|u_0|}{\sqrt{u_0^2 + 4t^2}} K \bigg( \frac{2t}{\sqrt{u_0^2 + 4t^2}} \bigg) .
\end{eqnarray}

The charge of a domain wall is equal to the sum of the charges of two `ends' of which it consists~\cite{brzezicki20}. For example, for the soliton in Figure~\ref{dptplot}(a), which consists of two consecutive sites with $-u_0$, the charge is
\begin{eqnarray}
Q_{\mathrm{sol}} \approx \frac{q_{\mathrm{e}}}{2} \left[ 1 - \zeta ( u_0 ) \right] .
\end{eqnarray}
For the antisoliton in Figure~\ref{dptplot}(a), which consists of two consecutive sites with $+u_0$, the charge is
\begin{eqnarray}
Q_{\mathrm{antisol}} \approx - \frac{q_{\mathrm{e}}}{2} \left[ 1 - \zeta ( u_0 ) \right] .
\end{eqnarray}


\begin{thebibliography}{99}
\bibitem{continuumnote}
If we made a different choice, e.g. $k \rightarrow (\pi / a) + \hat{p}/\hbar$, then the continuum Hamiltonian would be different by a unitary transformation $\sigma_z \hat{H} \sigma_z$ resulting in solitons with the same energy for the same profile of $u(x)$, but wave functions rotated as $\sigma_z \psi$.
\bibitem{smoothnote}
We present numerical data for atomically-smooth solitons~(\ref{smoothsol}) because the data for unit-cell-smooth solitons~(\ref{unitsmoothsol}) is qualitatively the same (for both soliton charge and robustness to disorder).
\end{thebibliography}
\end{document}